\documentclass[fleqn,usenatbib]{mnras}

\usepackage{newtxtext,newtxmath}

\usepackage[T1]{fontenc}
\usepackage{ae,aecompl}

\usepackage{graphicx}	        
\usepackage{amsmath}	        
\usepackage[]{algorithm2e}      

\makeatletter
\def\@printed{
    \qquad\qquad\qquad }
\gdef\@journal{\hfill\@printed}
\def\bsp{}
\def\@oddfoot{}
 \def\@evenhead{}
 \def\@evenfoot{}
\def\ps@titlepage{\let\@mkboth\@gobbletwo
 \def\@oddhead{\footnotesize\@journal}
 \def\@oddfoot{\hfil}
 \def\@evenhead{\hfill}
 \def\@evenfoot{\hfil}
 \def\sectionmark##1{}
 \def\subsectionmark##1{}}
\def\abstract{\if@twocolumn
   \start@SFBbox\@abstract
 \else
   \@abstract
 \fi}
\def\endabstract{\if@twocolumn
   \endlist\finish@SFBbox
 \else
  \endlist
 \fi}
\def\@abstract{\list{}{%
    \listparindent\realparindent
    \itemindent\z@
    \labelwidth\z@ \labelsep\z@
    \leftmargin 1.5pc\rightmargin 1.5pc
    \parsep 0pt plus 2pt}\item[]%
    \reset@font\normalsize{\bf ABSTRACT}\\\reset@font\large
}
\def\@maketitle{\newpage
 \vspace*{7pt}
 {\raggedright \sloppy
  {\reset@font\huge \bf \@title \par}
  \vskip 23pt
  {\reset@font\LARGE
   \begin{tabular}[t]{@{}l@{}}\let\\=\author@nextline\@author
   \end{tabular}
   \par}
  \vskip 22pt
 }
 \par\noindent
 {\reset@font\small \par}
 \vskip 22pt
}
\makeatother

\title[PINNs in hydrodynamic simulations]{Physics-informed neural networks in the recreation of hydrodynamic simulations from dark matter}

\author[Dai, Z. et al.]{
Zhenyu Dai$^{1,2}$\thanks{E-mail: \href{mailto:zdai3@uh.edu}{zdai3@uh.edu}},
Ben Moews$^{3,4,5,6}$,
Ricardo Vilalta$^{1,2}$,
Romeel Dav\'e$^{7,8,9}$
\\
$^{1}$Department of Computer Science, University of Houston, 3551 Cullen Blvd, Houston TX, 77204-3010, USA\\
$^{2}$Department of Physics, University of Houston, 3507 Cullen Blvd, Houston TX, 77204-5005, USA\\
$^{3}$Business School, University of Edinburgh, 29 Buccleuch Pl, Edinburgh, EH8 9JS, UK\\
$^{4}$Center for Statistics, University of Edinburgh, Peter Guthrie Tait Rd, Edinburgh, EH9 3FD, UK\\
$^{5}$McWilliams Center for Cosmology, Carnegie Mellon University, Hamerschlag Dr, Pittsburgh, PA 15213, USA\\
$^{6}$Pittsburgh Supercomputing Center, 300 South Craig St, Pittsburgh, PA 15213, USA\\
$^{7}$Institute for Astronomy, University of Edinburgh, Royal Observatory, Edinburgh EH9 3HJ, UK\\
$^{8}$Department of Physics and Astronomy, University of the Western Cape, Bellville, 7535, South Africa\\
$^{9}$South African Astronomical Observatories, Observatory, Cape Town 7925, South Africa\\
}

\newcommand{\fhat}{\hat{f}}
\newcommand{\N}{\textrm{NN}_{\theta}}

\newcommand{\simba}{{\sc Simba}\xspace}
\newcommand{\illustris}{{\sc Illustris}\xspace}
\newcommand{\mufasa}{{\sc Mufasa}\xspace}

\date{Accepted XXX. Received YYY; in original form ZZZ}

\pubyear{2023}

\begin{document}
\label{firstpage}
\pagerange{\pageref{firstpage}--\pageref{lastpage}}
\maketitle

\begin{abstract}
Physics-informed neural networks have emerged as a coherent framework for building predictive models that combine statistical patterns with domain knowledge. The underlying notion is to enrich the optimization loss function with known relationships to constrain the space of possible solutions. Hydrodynamic simulations are a core constituent of modern cosmology, while the required computations are both expensive and time-consuming. At the same time, the comparatively fast simulation of dark matter requires fewer resources, which has led to the emergence of machine learning algorithms for baryon inpainting as an active area of research; here, recreating the scatter found in hydrodynamic simulations is an ongoing challenge. This paper presents the first application of physics-informed neural networks to baryon inpainting by combining advances in neural network architectures with physical constraints, injecting theory on baryon conversion efficiency into the model loss function. We also introduce a punitive prediction comparison based on the Kullback-Leibler divergence, which enforces scatter reproduction. By simultaneously extracting the complete set of baryonic properties for the \textsc{Simba} suite of cosmological simulations, our results demonstrate improved accuracy of baryonic predictions based on dark matter halo properties and successful recovery of the fundamental metallicity relation, and retrieve scatter that traces the target simulation's distribution. 
\end{abstract}

\begin{keywords}
galaxies: evolution -- galaxies: haloes -- methods: analytical -- methods: statistical
\end{keywords}

\raggedbottom

\section{Introduction}
\label{sec:introduction}

The $\Lambda$CDM model, coined the standard model of cosmology due to its widespread adoption and explanatory power, plays a crucial role in modern cosmology and astrophysics. Galaxy formation and evolution occur within virialized structures resulting from density perturbations in the early Universe, subjected to gravitational collapse \citep{Frenk2012}. This leads to large-scale structure in the form of a cosmic web evolving from a somewhat smooth starting point, with dark matter halos as gravitationally bound overdensities of the postulated main contributor to the matter content of the Universe.

As galaxies form through the condensation of cooling gas within these halos, their resulting baryonic properties share a natural relationship with the dark matter accumulations they live in \citep[see, for example,][]{Rees1977, Blumenthal1984}. As dark matter distributions are often sufficient to extract cosmological parameters of interest, $N$-body simulations restricted to simulating dark matter particles or grids through gravitational interactions are common in cosmology \citep{Springel2005, Boylan-Kolchin2009, Klypin2011, Riebe2013, Potter2017}. Options in this regard include direct numerical integration and the inclusion of a scale factor to model the expansion of the Universe via general-relativistic effects. These simulations are computationally cheap and fast to run compared to more complex alternatives \citep{Efstathiou1985}.

In contrast to the success of large-scale structure, modeling galaxies through $N$-body simulations has historically been more difficult, as baryonic physics plays a vital role through nonlinear and dissipative processes at this level of granularity \citep{Somerville2015}. At the same time, the relevance of the baryonic properties of galaxies is evident by the reason dark matter bears its name; direct observations in the electromagnetic spectrum rely on baryonic processes emitting such radiation. Methods to include the luminous baryonic matter in our simulations are thus required to enable comparisons to these observations in the first place.

Multiple avenues to circumvent these limitations have been developed. These include abundance matching, which connects the halo mass to a range of baryonic properties through the stellar mass. However, this traditionally requires the assumption of the preservation of rank ordering in mass \citep[see, for example, descriptions in][]{Wetzel2010, Campbell2018}, although this has been superseded by an earlier shift towards the inclusion of intrinsic scatter as well as a preference for ordering by velocity instead of mass \citep{Reddick2013}. Recent models, however, do not solely rely on mass as the halo proxy but instead use a combination of factors and introduce scatter for a non-monotonic relation \citep{Lehmann2016, Stiskalek2022}. As an alternative to the rank ordering of all galaxies without separating central and satellite galaxies, halo occupancy distribution modeling makes assumptions about the satellite distributions and uses halo mass functions as an input \citep{Berlind2002}.

The penultimate step in this evolution of complexity are semi-analytic models, which incur a heavier computational burden as a trade-off for using a complete physical framework. The issue these models face is the number of free parameters for which constraints have to be found \citep[see, for example,][]{Somerville1999, Lu2014}. These include, but are not limited to, \textsc{GALFORM} as presented in \citet{Cole2000} and \citet{Baugh2018}, \textsc{GalICS} and \textsc{GalICS 2.0} \citep[see][]{Hatton2003, Cattaneo2017}, and \textsc{SAGE} \citep{Croton2016}. Simpler analytic formalisms for the evolution of baryonic galaxy properties exist, for example, the bathtub model by \citet{Bouche2010} and the reservoir model by \citet{Krumholz2012}, as well as the equilibrium model \citep{Dave2012, Saintonge2013, Mitra2015, Mitra2017}.

Hydrodynamic simulations, which require the most computational resources and time, are generally considered the gold standard of an ab initio approach to galaxy formation and evolution \citep{Vogelsberger2020}. Particle-based and grid-based methods exist based on the foundation that baryonic matter can be modeled by treating gas as an ideal fluid. The former rely on computations of discrete masses, or particles, while the latter splits the simulation volume into discrete spaces \citep{Dolag2008, Somerville2015}. Influential recent hydrodynamic simulations include, for example, \textsc{Illustris} and \textsc{IllustrisTNG} \citep[see][]{Genel2014, Pillepich2018}, \textsc{EAGLE} as introduced in \citet{Schaye2015}, and \textsc{HorizonAGN} by \citet{Dubois2016}, as well as \textsc{Mufasa} and its successor \simba, the latter of which is used for the lion's share of our experiments and is described in more detail in Section~\ref{sec:simba} \citep{Dave2016, Dave2019}.

In recent years, the physical sciences have experienced an exponentially rising interest in applying machine learning algorithms \citep{Carleo2019}. These developments include the acceleration of hydrodynamic simulations of galaxy formation and evolution, with \citet{Kamdar2016} providing one of the first examples by predicting baryonic properties in \textsc{Illustris} using extremely randomized trees, commonly abbreviated as `extra trees', an ensemble model based on decision trees. This heralded the dominance of tree-based ensembles in related research, including \citet{Agarwal2018} for \textsc{Mufasa} and \citet{Lovell2021} for \textsc{EAGLE}, as well as \citet{Jo2019}, \citet{McGibbon2022}, and \citet{deSanti2022}, for \textsc{IllustrisTNG}.

While these works apply tree-based ensembles, \citet{deSanti2022} also compare other algorithms such as $k$-nearest neighbors, light gradient-boosting machines, and feed-forward neural networks, and combine them with a linear regressor for improved predictions. Similarly, \citet{deAndres2022} use random forests, feed-forward neural networks, and natural and extreme gradient boosting on \textsc{The Three Hundred} data described by \citet{Cui2018}, and provide an example of favoring boosting over tree-based methods. In this case, extreme gradient boosting is reported to reach the best accuracy, while natural boosting retrieves the most scatter as a probabilistic regressor.

\citet{Jespersen2022} apply graph neural networks to \textsc{IllustrisTNG}. However, they only use the dark matter version and compute baryonic properties with a semi-analytic model, effectively making the predictor emulate such a model instead of a hydrodynamic simulation. Other works exist that focus on alternative algorithms or hybrid approaches. \citet{Moster2021}, for example, apply wide and deep neural networks (WDNN) and reinforcement learning. However, the baryonic properties are calculated not through a hydrodynamic simulation but \textsc{emerge}, an empirical model that statistically links properties from surveys instead of simulating baryonic physics \citep{Moster2018}. The prediction problem is flipped by \citet{vonMarttens2022}, who retrieve dark matter halos from their baryonic properties in \textsc{IllustrisTNG} instead.

\citet{Moews2021}, on the other hand, extend the equilibrium model by incorporating largest-progenitor merger trees\footnote{This might be a constant source of frustration for some readers of papers on these topics, as decision trees and merger trees are, apart from technically both having a tree-like structure, two entirely different concepts.}, and combine it with extra trees into a hybrid approach. One disadvantage of such analytic models is that they predict a limited set of properties. They can, however, be looped into both the training and prediction steps of machine learning algorithms by first predicting a limited set of properties with the analytic model and then using these predictions together with halo properties to predict the complete set of baryonics. Related work on hybrid models is done by \citet{Hearin2020}, who generate mock catalogs by combining empirical and semi-analytic models. This leads to the weighted Monte Carlo sampling of baseline catalogs to improve statistical realism.

\citet{Stiskalek2022}, similarly to \citet{Moster2021}, also use a WDNN approach for \textsc{IllustrisTNG} and \textsc{HorizonAGN}, and include a comparison with extra trees, but with a particular focus on reproducing the intrinsic scatter of the galaxy-halo connection. The reason is a well-known failure of commonly used machine learning algorithms when applied to hydrodynamic simulations to retrieve the expected scatter. The prediction of a Gaussian probability distribution for estimating targets does, of course, impose a constraint on the scatter. Still, probabilistic approaches to scatter reproduction are commonly encountered features in the literature \citep[see, for example,][]{Lehmann2016, Desmond2017, Mitra2017, Cao2020}.

Machine learning methods that rely on standard metrics like the mean squared error (MSE) face certain limitations. The dominant drawback for this application area is that these models aim to recreate the provided data without knowledge of the underlying physical theory. Some prior work tries to solve this issue; for example, the mentioned reports on weighted sampling through baseline libraries by \citet{Hearin2020} and pre-prediction of baryonic subsets to aid the machine learning model \citep{Moews2021}. Another more direct way is injecting theory directly into the learning process of neural network architectures, with such models coined physics-informed neural networks (PINN) \citep{Raissi2019}.

Originally developed for the finding of solutions for partial differential equations, PINNs are rooted in older research on neural networks for ordinary and partial differential equations \citep[see, for example,][]{Dissanayake1994, Lagaris1998}. In such models, knowledge represented as suitable equations offers domain constraints. Various approaches exist that include physical equations into the loss function while replacing a lengthy simulation with a machine-learning model has become common practice in many scientific endeavors \citep{Deiana2022}. The line of thinking behind this approach is simple; instead of evaluating the model solely on prediction accuracy through metrics such as the mean squared error, additional loss components can enforce compliance with additional parameter relationships. For a more in-depth overview of this rapidly expanding area of physics-driven deep learning, we refer interested readers to the reviews by \citet{Karnadiakis2021} and \citet{Cuomo2022}.

PINNs are not entirely alien to the field of astrophysics in general. Recent work by \citet{Mishra2021} targets the simulation of radiative transfer by minimizing the residual of the underlying transfer equations, while \citet{Martin2022} bypasses inefficiencies of gravity models to learn representations of small-body gravity fields directly. Other recent examples include \citet{Branca2022} on solutions for interstellar medium chemistry and the finding of quasinormal modes of nonrotating black holes \citep{Cornell2022}. The development and application of these architectures offer a powerful way to add knowledge to observational evidence to enforce adherence to theoretical models in training the machine learning algorithm.

In this paper, we present the first way to incorporate the concept of PINNs into the active research field of the completion of dark matter-only information in cosmological simulations with baryonic properties. To achieve this goal, we modify the training process of a deep learning model through two different novel extensions of the loss function. The first part is based on the standard approach of injecting physical theory into these models and adds the stellar-to-halo mass relation (SHMR) as described by \citet{Moster2010} into the training process. This double power law is subsequently used by \citet{Moster2018} to parameterize the instantaneous baryon conversion efficiency, meaning the efficiency with which gas is transformed into stars, and provides a theory-oriented constraint.

While this relation can, of course, be argued to be an empirical prescription rather than physical theory, an argument for being physics-informed is given through the baryon conversion efficiency, with higher $M_h$ values leading to the collection of more baryons. In detail, this link is more complicated, and although sufficient for an exploratory study on the benefit of these types of constraints, we discuss the limitations. In a broader sense, the description as physics-informed stems from the name of the model family itself. 

The second part, which presents another novel addition, forces the model to recreate the scatter of the underlying hydrodynamic simulation by including the Kullback-Leibler divergence (KLD) as an asymmetric distance measure between approximations of the probability distributions for training targets and associated model predictions \citep{Kullback1951, Ferdosi2011}. This differs in its motivation from the inclusion of the SHMR from \citet{Moster2010}, as it targets the recreation of scatter and not an improvement in mean accuracy. In doing so, we show that our extensions of existing machine learning approaches in baryon inpainting from dark matter halo properties are a powerful tool for modern cosmological simulations. 

Compared to the baseline model, our results demonstrate improvements in both predictive accuracy and the reproduction of scatter, and show suitable correlations between target variables and model predictions. The contribution pertains to the broader field of machine learning in astrophysics, including adding physical theory into predictive models and using distributional loss components.

The remainder of this paper is structured as follows. In Section~\ref{sec:methodology}, we provide an overview of our machine learning approach and data. Section~\ref{sec:simba} describes the \simba suite of cosmological simulations and our dataset. Section~\ref{sec:pinns} covers the functionality and justification of our baseline PINN model; Sections~\ref{sec:theory_loss} and~\ref{sec:kld_loss} introduce extensions of the loss function for the SHMR and a distribution comparison between predictions and target values, respectively. Section~\ref{sec:experiments} presents our experiments and their results. Specifically, Section~\ref{sec:shmr_fitting} explains the fitting of our theory constraint to \simba and the weighting of loss function components. Section~\ref{sec:predictions} provides the results for our predictions, and Section~\ref{sec:correlations} shows correlations for individual data point accuracy. Section~\ref{sec:discussion} discusses our findings, limitations of our approach, and follow-ups. Lastly, Section~\ref{sec:conclusion} provides our conclusions.

\section{Data and methodology}
\label{sec:methodology}

\subsection{Simulation data from the \simba suite}
\label{sec:simba}

The \simba simulation models the co-evolution of gas and dark matter within an expanding metric using the {\sc Gizmo} code~\citep[see][]{Hopkins2015}, which itself is based on the {\sc Gadget-2}~\citep{Springel2005}.  It employs the Meshless Finite Mass (MFM) hydrodynamics method, which marries the convenience of a mass-conserving particle-based code with the shock and instability-capturing advantages of a Riemann solver-based scheme.  

Many so-called sub-grid processes have been added to {\sc Gizmo} to model the formation and evolution of galaxies.  These include radiative cooling and photoionization heating, chemical enrichment from stellar evolution, the formation of stars and supermassive black holes, the energy release (`feedback') from supernovae and black hole accretion discs, and the growth and destruction of dust.  The complete model detailing these sub-grid prescriptions is described in \citet{Dave2016, Dave2019}.

\simba simulations begin in the linear regime at redshift $z=249$, and are evolved to $z=0$, meaning today.  151 snapshot outputs are stored at various redshifts along the way.  The main \simba run models a random cube of 147~Mpc (comoving) on a side, represented by 1024$^3$ gas elements and 1024$^3$ dark matter particles.  The minimum (adaptive) spatial resolution is 0.7~kpc.  The simulation assumes a {\it Planck}-concordant cosmology \citep[see][]{Planck2016} of $\Omega_m=0.3$, $\Lambda=0.7$, $H_0=68$~km/s/Mpc, $\sigma_8=0.82$, and $n_s=0.97$.  This results in a mass resolution of $1.8\times 10^7$ per gas element and $9.5\times 10^7$ per dark matter particle.

Each snapshot is analyzed using the {\sc Caesar} galaxy/halo catalog package.  For each halo identified within \simba using {\sc Gizmo}'s native 3D Friends-of-Friends (FoF) finder, {\sc Caesar} identifies galaxies as collections of stars and dense gas via a 6D FoF algorithm.  The most massive galaxy within a halo is defined as the central, and the others are satellites.  A large range of physical and photometric properties are computed for each halo and galaxy.  For this work, the key galaxy quantities are the stellar mass ($M_*$), star formation rate (SFR), SFR-weighted gas-phase metallicity ($Z$), neutral hydrogen mass ($M_\mathrm{HI}$), molecular hydrogen mass ($M_\mathrm{H2}$), and central supermassive black hole mass ($M_\mathrm{BH}$). 

These properties are obtained by summing the relevant particles in each galaxy.  For halos, the relevant quantities are the total mass ($M_h$), the dark matter half-mass radius ($r_h$), and the dark matter velocity dispersion ($\sigma_h$).  The catalogs are stored as HDF5 files, and {\sc Caesar} provides a simple yet powerful Python-based access interface. The \simba snapshots and catalogs are all publicly available online for use by the scientific community\footnote{\url{https://simba.roe.ac.uk}}.

Our work uses central galaxies from the \texttt{m100n1024\_151} version of the \simba main runs. Entries for which $M_\mathrm{BH} = 0$ are dropped to avoid zero-mass black holes distorting the distributions of predictions in our experiments, which is in line with related research covered in Section~\ref{sec:introduction}. This allows predictive models to focus on the recovery of accurate shapes for exploratory studies, while a sharp cutoff from seeding processes in hydrodynamical simulations complicates this in continuous prediction spaces. These considerations, as well as potential solutions, are further discussed in Section~\ref{sec:discussion}. 

Similarly, we restrict the range of included halo masses to  $11 \leq \log_{10}({M_{h}}) \leq 14$, following the same related research. Apart from these preprocessing steps, we refrain from any further alterations. While suitable data selections could lead to improved predictions, the goal of this paper is to maintain generalizability. For the model training and prediction processes, we apply min-max normalization to input and target variables to scale values within the same interval $[0, 1]$ and then revert predictions to their proper scales.

\subsection{Physics-informed neural network model}
\label{sec:pinns}

The concept of PINNs rests on the assumption that data-driven statistical learning can be enhanced with domain knowledge. This integration can be implemented using different frameworks, as covered in the overview of Section~\ref{sec:introduction} and related reviews \citep{Karnadiakis2021, Cuomo2022}. One approach is to generate more data specifically crafted to enforce domain knowledge into the model. While simple, this approach requires large amounts of additional data to cover a broad region of the input-variable space, effectively introducing observational biases into the inductive analysis \citep{Kashefi2021, Yang2019}. 

Another approach involves designing specific learning algorithms that embed knowledge into the learning architecture, for example, convolutional neural networks for images and speech \citep[see][]{LeCun1995}, graph neural networks \citep[see][]{Zhou2020, Wu2020}, and networks for Hamiltonian systems, among others \citep{Jin2020}. Implementing this approach is difficult because embedding physical laws within a neural network architecture is limited to simple processes. Complex processes require architectural designs that cannot be easily realized with current learning frameworks; even relatively simple processes need complex and elaborate designs. Regarding practical applications, a second complication of these purpose-designed models is creating a network specific to a given problem, making the transfer to new domain applications time-consuming.

The third approach we adopt here enriches the loss function to explicitly incorporate constraints, usually as partial differential equations \citep{Raissi2019, Lagaris1998}. The benefit is a decoupling between the machine learning strategy and the embedded knowledge; the corresponding framework has broader applicability since the learning architecture is built independently of the underlying physical laws. Normally, the loss function is extended to include a measure of the accuracy of each prediction and the degree of alignment with a physical constraint. 

The ability of PINNs to generalize well goes beyond approximation theorems. The representational power of neural networks is well known; under limited assumptions, any continuous function can be approximated to an arbitrarily close fit using a neural network with a finite number of hidden nodes and one hidden layer \citep{Cybenko1989, Yarotsky2017}. It should be noted that representational power is not equivalent to generalization power. Adding domain knowledge to the loss function improves the generalizability of PINNs by reducing the bias component of error while keeping the variance component under control \citep{Hastie2009}.

In traditional PINNs, the final loss, $\mathcal{L}_f$, is a weighted combination of two losses; one is the data-driven empirical loss, $\mathcal{L}_s$, and the other is the domain-knowledge constraint, $\mathcal{L}_k$, with respective weights $w_s$ and $w_k$, so that
\begin{eqnarray}
\label{eq:final_loss}
\mathcal{L}_f = w_s \mathcal{L}_s + w_k \mathcal{L}_k.
\label{eq:theory_loss}
\end{eqnarray}
The first term is the conventional loss obtained in traditional neural networks through methods such as the squared difference between predictions and target values. Training a network $\N$ that aims to find a near-optimal parameter vector $\theta$, for example, through gradient descent, yields a surrogate for a solution to processes such as complex simulations. The corresponding output $\fhat(x)$, where $x$ is a feature vector, serves to adjust the weights during training by looking at a loss function $\mathcal{L}_s(\fhat(x), y)$. Here, $y$ is the true response variable, or the target value from the underlying hydrodynamic simulation in our case, and $\fhat(x)$ is our model-provided estimate.

The additional term, $\mathcal{L}_k$, adds domain constraints, usually as differential equations. Specifically, the term captures the partial differential residuals as covered in Section~\ref{sec:theory_loss}. In our study, the inputs to the neural network are $x:= [M_h,r_h,\sigma_h]$ corresponding to dark matter halo mass at present, dark matter half-mass radius, and dark matter halo velocity dispersion, respectively. 

The outputs are $y:=[M_*, \mathrm{SFR}, Z, M_\mathrm{HI}, M_\mathrm{H2}, M_\mathrm{BH}]$ corresponding to stellar mass, star formation rate, metallicity, neutral and molecular hydrogen masses, and black hole mass, respectively. During training, we aim to minimize the residual sum of squares,

\begin{eqnarray}
\mathcal{L}_s = \frac{1}{N} \sum_{i = 1}^N \sum_{j = 1}^M (y^{(j)}_{i} - \fhat^{(j)}(x_{i}))^2.
\label{eq:mse_loss}
\end{eqnarray}

Here, $N$ is the number of examples in the training data, while $M$ is the respective target space subject to prediction.

As the subsequent distribution-based loss is calculated over each epoch's output for the training data, we take the arithmetic mean of the residual sum of squares for each training iteration.
Domain knowledge minimizes a different loss function, $\mathcal{L}_k$, forcing the final model to obey the constraint from physical knowledge. Here, we assume points $\{x\}$ sampled across the entire input space. In the next section, we explain the domain loss in detail. With the defined loss functions, the neural network $\N$ is trained to obtain parameters $\theta$ using efficient optimization methods, such as gradient descent. The weights $w_s$ and $w_k$ enable different contributions to the final loss function and can be tuned automatically as part of the optimization process. 
 
\subsection{Inclusion of baryon conversion efficiency}
\label{sec:theory_loss}

As the name suggests, the stellar-to-halo mass relation links the stellar mass of a given galaxy to the dark matter halo mass. Suitable parameterizations have been shown to reflect the galaxy mass function observed in the third data release of the Sloan Digital Sky Survey \citep[SDSS DR3; see][]{Panter2007} more closely, as it does not assume a constant SHMR. The instantaneous baryon conversion efficiency $\epsilon$ is the rate at which gas is transformed into stars. This efficiency can be described by an SHMR parameterized through a double power law model as shown by \citet{Moster2010}.

The latter introduces a parameterization that follows observations by avoiding a surplus of galaxies at low and high masses. For two slopes $\beta$ and $\gamma$ that are used to determine the decrease in efficiency at lower and higher masses, respectively, the parameterization takes the form
\begin{eqnarray}
    \epsilon(M, z = 0) = 2 \epsilon_N \left( \left(\frac{M_h}{M_1} \right)^{-\beta} + \left( \frac{M_h}{M_1} \right)^{\gamma}\right)^{-1}.
\label{eq:shmr}
\end{eqnarray}
Here, $\epsilon_N$ is the normalization, while $M_1$ denotes the characteristic mass at which the respective efficiency is the same as its normalization. \citet{Moster2018} use this to parameterize the instantaneous baryon conversion efficiency, showing that peak conversion efficiency takes place at halo masses similar to the characteristic mass,
\begin{eqnarray}
    M_{\mathrm{max}} = M_1 \left( \frac{\beta}{\gamma} \right)^{(\beta + \gamma)^{-1}},
\end{eqnarray}
with the general assumption of $\beta, \gamma > 0$. The integrated baryon conversion efficiency is dependent on redshift \citep[see][]{Moster2013}, and the mentioned work allows for parameters of the instantaneous efficiency to vary, with
\begin{eqnarray}
\begin{aligned}
    \log_{10} M_1 (z) &= M_0 + M_z (1 - (z + 1)^{-1})\\
    &= M_0 + M_z \left( \frac{z}{z + 1} \right),
\end{aligned}
\label{eq:shmr_parameters_1}
\end{eqnarray}
and with the normalization and slopes given by
\begin{eqnarray}
\begin{aligned}
    \epsilon_N(z) &= \epsilon_0 + \epsilon_z (1 - (z + 1)^{-1}) = \epsilon_0 + \epsilon_z \left( \frac{z}{z + 1} \right),\\
    \beta(z) &= \beta_0 + \beta_z (1 - (z + 1)^{-1}) = \beta_0 + \beta_z \left( \frac{z}{z + 1} \right),\\
    \gamma(z) &= \gamma_0.
\end{aligned}
\label{eq:shmr_parameters_2}
\end{eqnarray}
As we operate at $z = 0$, these considerations are simplified and need only be optimized for single values. In Section~\ref{sec:shmr_fitting}, we will cover this optimization for \simba data using maximum likelihood estimation, with ranges provided by prior research on this parameterization. This view on the SHMR can then be included, for a given dataset size of $N$, into the loss function of Eq.~\ref{eq:theory_loss} as
\begin{eqnarray}
\mathcal{L}_k = \sum_{i = 1}^N \left( \frac{\hat{M_*}}{M_h}-2 \epsilon_N \left( \left( \frac{M_h}{M_1} \right) ^{-\beta} + \left( \frac{M_h}{M_1} \right)^{\gamma} \right)^{-1} \right)^2.
\label{eq:shmr_loss}
\end{eqnarray}
This injection of domain knowledge provides an additional constraint for the model, which subsequent experiments show to benefit the model in recovering the mean relation better.

Including the $M_*$-$M_h$ relationship used in this study is physics-informed through the link to the baryon conversion efficiency, which is given through the fact that bigger dark matter halos are able to collect more baryons, thus forming more stars. This is, of course, a simplified treatment, but sufficient for exploratory work, and drawbacks are further discussed in Section~\ref{sec:discussion}.

\subsection{Constraints from predictive distributions}
\label{sec:kld_loss}

Reproducing the scatter found in hydrodynamic simulations when predicting baryonic properties based on dark matter halos is a known challenge in the literature \citep{Cui2018, Stiskalek2022}. As described in Section~\ref{sec:introduction}, probabilistic approaches relying on parameterized distribution families are common. In this work, we target the reproduction based on the training data distribution directly by introducing a second extension to the standard loss function in Eq.~\ref{eq:mse_loss} (see Section~\ref{sec:pinns}).

The Kullback-Leibler divergence (KLD) is a statistical distance measure to assess the difference between two distributions. Introduced by \citet{Kullback1951}, it has found various applications in astrophysics in recent years \citep[see, for example,][]{Ben-David2015, Hee2016, Moews2019, Nicola2019}. For a given reference distribution $P$ and proposal distribution $Q$, the KLD can be written as
\begin{eqnarray}
D_{\text{KL}}(P || Q) = \sum_{x \in \chi} P(x) \text{log} \frac{P(x)}{Q(x)},
\label{eq:kld}
\end{eqnarray}
or, correspondingly, with an integral for absolutely continuous probability distributions. One important point is that the KLD is not a distance metric due to its status as an asymmetric difference measure. This means that 
\begin{eqnarray}
D_{\text{KL}}(P || Q) \neq D_{\text{KL}}(Q || P),
\end{eqnarray}
as it calculates a directional information loss when approximating $P$ via $Q$. It also does not satisfy the triangle equality, 
\begin{eqnarray}
d(a,c)\leq d(a,b) + d(b,c),
\end{eqnarray}
with points $\{a, b, c\} \in M$ for a given metric space $M$. Given the above, the KLD is applicable only when a `true' reference distribution is used. Fortunately, this is the case here, as we want to calculate the difference between the distributions of model predictions and their respective targets.

Following \citet{Fussell2019}, who propose the incorporation of the KLD into the loss function in the context of generative modeling -- although without implementing the proposal due to conflicting success metrics -- we extend Eq.~\ref{eq:final_loss} to
\begin{eqnarray}
\mathcal{L}_{f} = w_s \mathcal{L}_{s} + w_k \mathcal{L}_k + w_{\mathrm{KL}} \mathcal{L}_{\mathrm{KL}}.
\label{eq:hybrid_loss}
\end{eqnarray}
Here, the KLD is part of the overall loss function through a normal assumption placed on the target and prediction distributions, as the loss needs to remain easily differentiable for error backpropagation during training. This means that
\begin{eqnarray}
\mathcal{L}_{\mathrm{KL}} = D_{\mathrm{KL}} \big( \mathcal{N}(\mu, \sigma) || \mathcal{N}(\hat{\mu}, \hat{\sigma}) \big),
\label{eq:kld_loss}
\end{eqnarray}
for target mean and variance, $\mu$ and $\sigma$, and corresponding predictions, $\hat{\mu}$ and $\hat{\sigma}$. In doing so, and as described in the overview in Section~\ref{sec:introduction}, we impose an additional constraint on the learning process by forcing the model to recreate the approximate distribution of the underlying training data. This extension focuses on the more accurate reproduction of the scatter found in the underlying simulation, not on the improvement of the mean accuracy. In later sections, we will see how this avoids an underprediction of tails in the scatter of the baryonic properties of interest.

\begin{figure}
\includegraphics[width=\columnwidth]{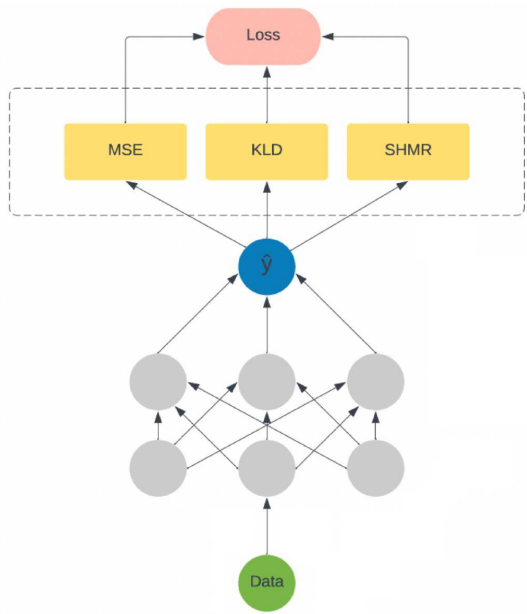}
\caption{Illustration of the hybrid model. The loss function used in this framework is a combination of the mean squared error, the Kullback-Leibler divergence, and the stellar-to-halo mass relationship.}
\label{fig:figure_1}
\end{figure}

Figure~\ref{fig:figure_1} provides a schematic for the complete model incorporating both the SHMR from Section~\ref{sec:theory_loss} and the KLD, which we dub the `Hybrid' model going forward. The lower part of the figure shows the neural network model, going from input data containing information from a set of dark matter halo properties to baryonic predictions, $\hat{y}$, as the model output. The components of the loss function are encased by a dashed line, listing the mean squared error, the KLD introduced in this section, and the SHMR. The respective weights of these components can be selected based on the predictive performance of the neural network, which we cover as part of our experiments in Section~\ref{sec:shmr_fitting}.

The normal assumption made for the respective distributions is, of course, not without fault. In particular, it remains to be seen in the results whether this assumption can lead to a decrease in mean predictive accuracy for sufficiently non-normal distributions. However, it provides an approximation that is differentiable for the purpose of backpropagation during training and is computationally cheap enough to result in reasonable training times on small numbers of GPUs without requiring access to stacks in large-scale supercomputing solutions. We reserve part of the discussion in Section~\ref{sec:discussion} to list possible alternatives in related follow-up research.

\section{Experiments and results}
\label{sec:experiments}

\subsection{The stellar-to-halo mass relation in \simba}
\label{sec:shmr_fitting}

We first train a simple feed-forward neural network, a multilayer perceptron (MLP), using only the MSE in the loss function, and then use this baseline model to predict all six parameters. In order to efficiently explore and optimize the hyperparameter space, we adopt a sequential model-based optimization (SMBO) approach. The SMBO algorithm iteratively updates the surrogate model, explores the hyperparameter space by sampling new configurations, evaluates their performance using the objective function, and refines the search process based on the observed results, allowing for a reasonably fast and continuous optimization. For this, we use the Optuna hyperparameter optimization framework to determine the optimal number of layers and neurons in the hidden layers~\citet{optuna_2019}. The search space was set to be between 3--10 layers and 30 to 100 neurons per layer, following common choices for similarly structured problems.

After invoking the optimization module, the best configuration displayed nine hidden layers with 80 artificial neurons, each with a ReLU activation function, equally following current practise. In addition, we split off $10\%$ of the dataset as a validation set to perform a search on the weights, freely chosen in $[0,1]$, for the loss function components, $w = [w_s, w_k, w_\mathrm{KL}]$. These weights are optimized through a combination of grid-search Bayesian optimization and meta-heuristics using Optuna~\citep{optuna_2019}. They are optimized to minimize the overall loss on a validation set, averaged over ten runs. As in most optimization techniques, the trivial solution of assigning the least possible value to each weight is avoided by imposing a constraint on the weights, meaning the sum of the weights is a constant.

\begin{figure}
\includegraphics[width=\columnwidth]{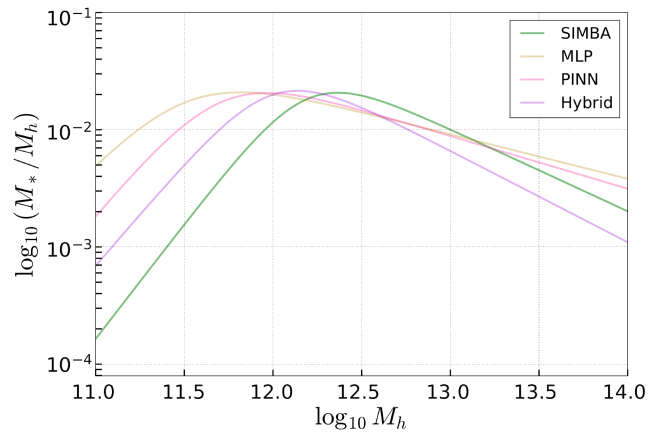}
\caption{Double power law fitting of model predictions for the baseline MLP (yellow), the PIUNN (red), and the hybrid PINN + KLD model (purple). The \simba data is shown in green.}
\label{fig:figure_2}
\end{figure}

\begin{figure*}
\includegraphics[width=\textwidth]{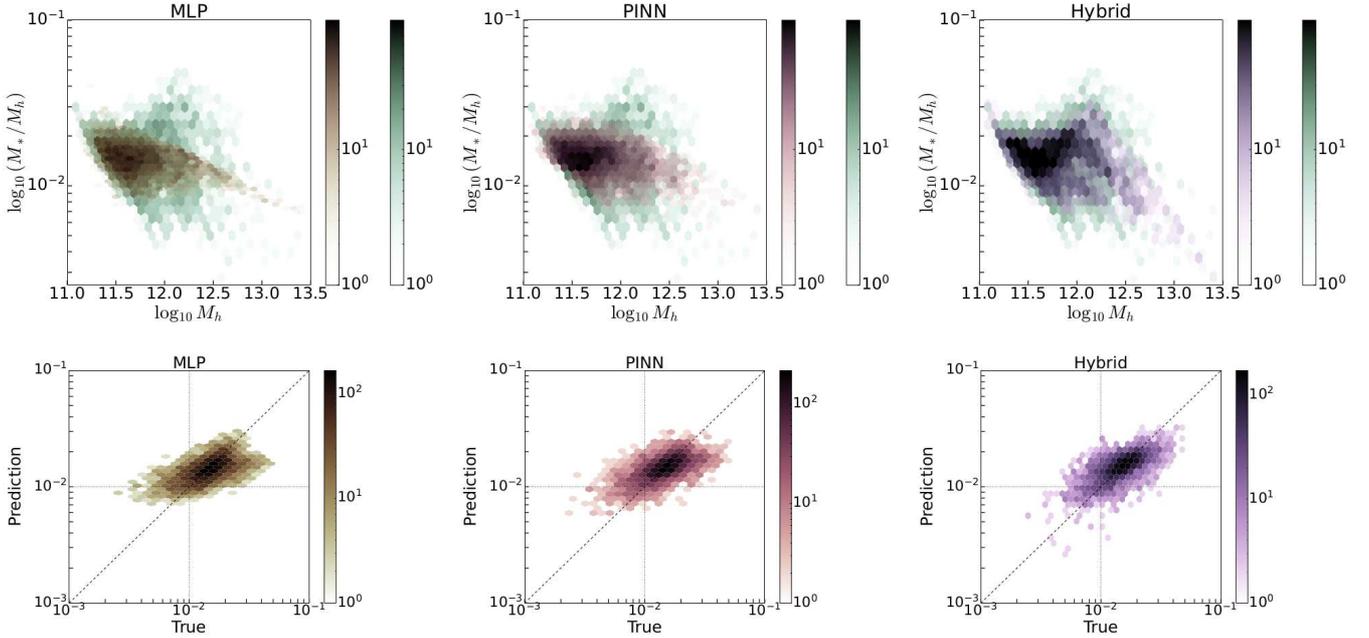}
\caption{Top: Hexagonal joint histograms of the predicted SHMR from different models and the true SHMR calculated from the \simba test set. From left to right, the panels show the results from the baseline MLP (yellow), the PINN (red), and the hybrid PINN + KLD model (purple). The \simba data is shown in green. Bottom panels: Hexagonal joint histograms of the predicted versus target SHMR for different models and the test set, with color coding as described above. The diagonal line indicates $x = y$ for the resulting correlation plots.}
\label{fig:figure_3}
\end{figure*}

While an MLE approach is infeasible due to the model retraining for each run, this allows us to gauge a suitable combination of weights. The best-fit combination is tested for 10-fold cross-validation using the MSE and the mean absolute percentage error (MAPE), with the results listed in Table~\ref{tab:table_2}. As Figure~\ref{fig:figure_2} and the top panels of Figure~\ref{fig:figure_3} show, this basic model can recover the basic pattern of the SHMR represented in Eq.~\ref{eq:shmr}. However, the pattern exhibits a diminished variance at lower and higher halo masses and is expectedly flat around the characteristic $M_{\mathrm{max}}$ value.

\begin{table}
\centering
    \caption{Optimization of best-fit loss function component weights using SMBO. The table lists the mean squared error and mean absolute percentage error for various loss function weight combinations in the hybrid model. The values in brackets list the associated standard deviations.}
    \begin{tabular}{l c c}
    \hline
    $(w_s,w_k,w_\mathrm{KL})$ & MSE & MAPE \\
    \hline
    $(1,0.6,1.3)$ & 0.02579 (0.0016) & 1.098 (0.0532) \\
    \hline
    \end{tabular}
    \label{tab:table_2}
\end{table}

As a result, the predictions show a flatter relation than 1:1 relative to the true distribution from \simba, indicating a bias.  We note that \simba's galaxy sample is limited in stellar mass, which results in a diagonal completeness limit in the SHMR. This leads to galaxies with a high $M_*/M_h$ ratio for their halo mass being preferentially included in the sample; such asymmetric completeness (known as Malmquist bias) is common in astrophysics, but can be difficult for machine learning algorithms to recover.

Our PINN, on the other hand, is constructed by encoding Eq.~\ref{eq:shmr} into the loss function as shown in Section~\ref{sec:theory_loss}. The parameters of the resulting loss component shown in Eq.~\ref{eq:shmr_loss} are estimated through maximum likelihood estimation (MLE). MLE is performed to optimize the parameters of Eq.~\ref{eq:shmr} and, accordingly, of the PINN part of the loss function in Eq.~\ref{eq:shmr_loss}. Here, the likelihood function is the density function regarded as a function of given parameters $\theta$,
\begin{eqnarray}
    \mathcal{L}(\theta) = \sum_{i = 1}^n f(x_i | \theta), \ \mathrm{with} \ \theta \in \Theta,
\end{eqnarray}
where $\Theta$ is the corresponding parameter space and $f(\cdot)$ is the probability density function. We can estimate the optimal combination of parameter values of interest, $\hat{\theta}$, as
\begin{eqnarray}
    \hat{\theta} = \arg \max_{\theta \in \Theta} \mathcal{L} (\theta | x),
\end{eqnarray}
which results in a choice of parameter values that make the observed data the most probable. Different values in a wide range have been reported on Eq.~\ref{eq:shmr} \citep{Behroozi2019, Kravtsov2018, Shankar2017}. Instead of directly applying these numbers, they become the initial guess of our estimator, and the range of values is considered a constraint. The best-fit parameters for the test set are $\epsilon_N = 0.02, M_1=12.19, \beta=1.1, \gamma=0.78$. With this extra constraint, the predicted SHMR pattern better recovers the curvature around $M_{\mathrm{max}}$ in Figure~\ref{fig:figure_3}. Eq.~\ref{eq:shmr}, however, as a double power law function, this additional constraint cannot introduce the scatter found in the cosmological simulation and still displays a slightly biased recovery of the true SHMR relation.

To retrieve the scatter, we need a way to provide information as part of the loss function, effectively forcing the model to recreate it. For this, we extend the loss function with an additional component using the KLD, which we introduced in Section~\ref{sec:kld_loss}. This distributional divergence measurement has a history of being used in various cosmological applications, as discussed in Section~\ref{sec:introduction}, including the study of properties and auxiliary observational data on baryonic physics \citep{Yasin2022}. The resulting predictions from this hybrid scheme track the respective target values more faithfully and with less bias.

We calculate the mean and covariance matrix of the training data and construct a multivariate Gaussian distribution in six dimensions, one for each prediction target property. By including the KLD-based loss in Eq.~\ref{eq:kld} into the overall loss function, we create a hybrid model combining the PINN with a KLD measurement as written in Eq.~\ref{eq:hybrid_loss}. Table~\ref{tab:table_1}, listing the MSE as well as the coefficient of determination ($R^2$) and the Pearson correlation coefficient ($\rho$), demonstrates the benefit on the predictive power of these models.

\begin{table}  

\centering
\caption{Performance comparison between different models. The table lists the coefficient of determination ($R^2$) and Pearson's correlation coefficient ($\rho$) for different neural network models corresponding to Figure~\ref{fig:figure_2} and Figure~\ref{fig:figure_2}, as well as the best-fit relation following \citet{Moster2010}. The hybrid model is a combination of PINN with KLD in the loss.}
\begin{tabular}{l c c c}
\hline
Model & MSE & $R^2$ & $\rho$  \\
\hline
Moster relation & 0.545 & 0.730 & 0.855 \\
MLP & 0.024  & 0.821 & 0.906 \\
PINN & 0.023 & 0.827 & 0.910\\
Hybrid (PINN+KLD) & 0.020 & 0.848 & 0.921 \\
\hline
\end{tabular}
\label{tab:table_1}

\end{table}

We plot the SHMR relative to the halo mass in Figure~\ref{fig:figure_3} to provide a visual overview. This confirms that the hybrid approach between extra physical knowledge and distributional adherence outperforms both the baseline model and the PINN alone. The latter provides information on the mean trends of the SHMR, while the KLD helps the neural network to mimic the substantial scatter, which an exact equation does not provide. Compared to the baseline MLP, the figure shows that the distributional component helps predict the SHMR more accurately at lower halo masses while tracing the downward scatter at higher halo masses.

To better understand the correlation between targets and predictions, the bottom panels of Figure~\ref{fig:figure_3} show the SHMR values from \simba on the horizontal axis, with SHMR values from model predictions on the vertical axis. The result would follow the plot's diagonal line for a perfect model. The visuals, which indicates a better correlation for the PINN compared to the MLP, and for the hybrid model compared to both, are confirmed by the corresponding correlation measurements in Table~\ref{tab:table_1}. As discussed in Section~\ref{sec:kld_loss}, the normal assumption for the KLD constraint could be argued to enable a decrease in mean accuracy in some cases, and these preliminary results demonstrate the overall ability of the model to use the constraint in a non-detrimental manner. Apart from the SHMR, the KLD also improves the model's prediction ability on other baryonic properties, which we will cover, together with potential explanations for an improvement instead of no discernible difference in case of no negative effect of the normal assumption, in Sections~\ref{sec:predictions} and~\ref{sec:discussion}.

\subsection{Predicting baryonic properties from dark matter halos}
\label{sec:predictions}

\begin{figure}
\includegraphics[width=\columnwidth]{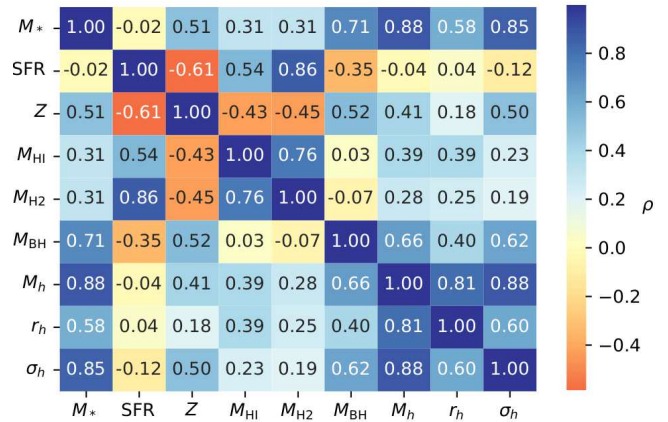}
\caption{Correlation matrix of input and output parameters. The color bar indicates the Pearson correlation coefficient values, calculated for variables in the \simba dataset used in the presented work.}
\label{fig:figure_4}
\end{figure}

After establishing the effect on SHMR prediction, we expand our analysis to the full set of six baryonic properties, $\{M_*, \mathrm{SFR}, Z, M_\mathrm{HI}, M_\mathrm{H2}, M_\mathrm{BH}\}$. To better understand the cause of these results, we calculate $\rho$ values between the entire set of available parameters in Figure~\ref{fig:figure_4}.

\begin{figure*}
\includegraphics[width=\textwidth]{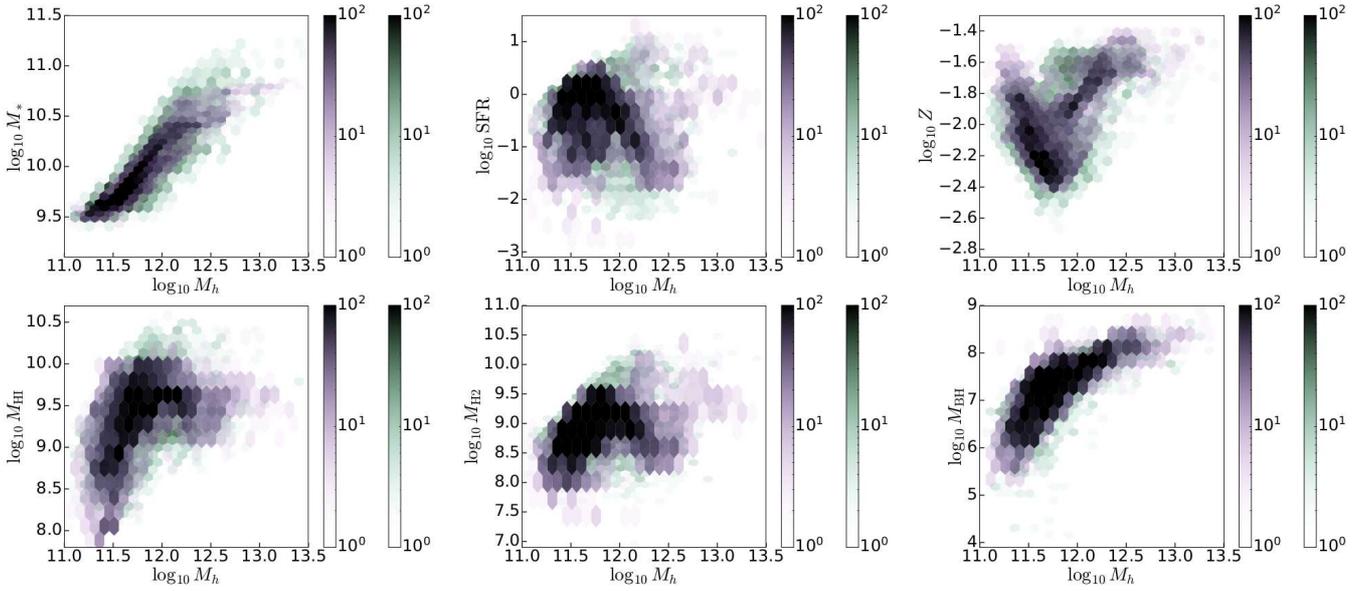}
\caption{Hexagonal joint histograms of target values and model predictions. The panels show, for the set of six baryonic properties of interest, plots against the dark matter halo mass. Model predictions are in purple, while the \simba data is in green.}
\label{fig:figure_5}
\end{figure*}

Figure~\ref{fig:figure_5}, where these variables are plotted against the halo mass, shows that each parameter is reasonably well-predicted upon visual inspection. In particular, the model excels for $M_*$ and $M_\mathrm{BH}$, while the performance on SFR and $M_\mathrm{H2}$ is subject to a scatter taper at higher halo masses. In the values of Figure~\ref{fig:figure_4}, we can see that the correlation between the dark matter halo properties used as inputs, meaning $\{M_\mathrm{h}, r_h, \sigma_h\}$, for these variables are considerably lower compared to the rest of the investigated properties. At the same time, good results are equally reflected in strong correlations for stellar and black hole masses. As covered in Section~\ref{sec:shmr_fitting}, this is aided by the additional SHMR constraint in the loss function.

\begin{figure}
\includegraphics[width=\columnwidth]{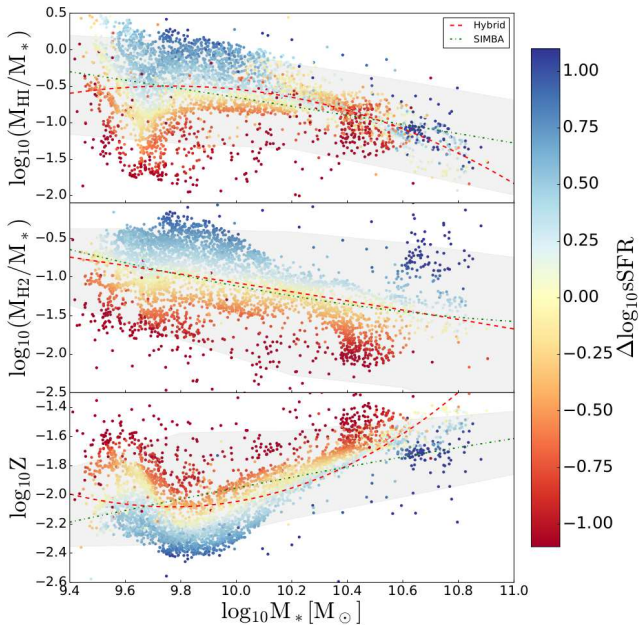}
\caption{Secondary correlations at $z = 0$. The panels show, from top to bottom, neutral hydrogen, molecular hydrogen, and metallicity as a function of the stellar mass. Mean scaling relations are drawn with green dot-dashed lines for \simba and red dashed lines for our hybrid model. Galaxies are colored by the distance from the mean $M_*-\mathrm{sSFR}$ relation.}
\label{fig:figure_6}
\end{figure}

Following this, we explore secondary correlations between different variables, analogous to a similar analysis performed by \citet{Agarwal2018}. Figure~\ref{fig:figure_4} indicates that galaxies with higher $\mathrm{SFR}$ exhibit positive correlations with $M_\mathrm{HI}$ and $M_\mathrm{H2}$, and a negative correlation with $Z$. To test whether our model correctly learns the split in the specific star formation rate,
\begin{eqnarray}
    \mathrm{sSFR} = \frac{\mathrm{SFR}}{M_*},
\end{eqnarray}
We plot these properties against the stellar mass and color data points using the distance in $\mathrm{sSFR}$ values from the mean $M_*-\mathrm{sSFR}$ relation, $\Delta \log_{10} \mathrm{sSFR}$, in Figure~\ref{fig:figure_6}. The color coding highlights the ML model's recovery of the second-parameter correlation through the sSFR, which is also seen in Simba. This means that the data points plotted in the figure are predictions of our model, not data points extracted from the simulation. The resulting mean scaling relations are drawn for model predictions and the corresponding data from the underlying \simba simulation, with the former tracing the results of \citet{Agarwal2018} and \citet{Dave2019}.

Both mean scaling relations are obtained by fourth-order polynomial fitting, and we provide the 6$^\mathrm{th}$ to 93$^\mathrm{th}$ percentile range for \simba data indicated by grey shading. In doing so, the figure provides the simulation distribution as a reference in addition to the color-coded model predictions. We use a bin size of 0.2 for the $\log_{10} M_*$ values in solar masses along the horizontal axis for the latter. The divergence between confidence intervals and the mean scaling relation for metallicity in \simba at lower stellar masses is an artifact of a small number of data points available at this range, but we include this left-hand interval to exemplify potential peculiarities encountered in such analyses.

As our approach is a prediction problem using approximations, it is not perfectly consistent, just like other works on machine learning for baryonic inpainting into dark matter halos. Here, the underprediction of scatter at higher stellar masses is notable for $M_\mathrm{H2}$ compared to \simba. At the same time, the relationships of $Z$, $M_\mathrm{HI}$, and $M_\mathrm{H2}$ to the sSFR are preserved around the mean scaling relations, although in a cleaner split than is the case in the target values provided by the \simba cosmological simulation suite.

While similar patterns can be found in \simba, the discrepancies in polynomial fits can be attributed to the wave-like nature common to neural network predictions. This can depend on the activation function and refers to the nonlinear change in the vertical distribution of data points in Figure~\ref{fig:figure_6}. While common features in these types of models, the present exploratory study demonstrates the viability of our approach and is discussed further in Section~\ref{sec:discussion}. MLP predictions without extensions to the loss function severely underpredict the scatter across the board at higher stellar masses. The overall distribution of predictions stays withing the \simba-indicated areas. The recovery of secondary correlations between SFR (or gas content), metallicity, and stellar mass, known as the fundamental metallicity relation, is an important success of this machine learning framework.

\subsection{Correlations for separate prediction targets}
\label{sec:correlations}

\begin{figure*}
\includegraphics[width=\textwidth]{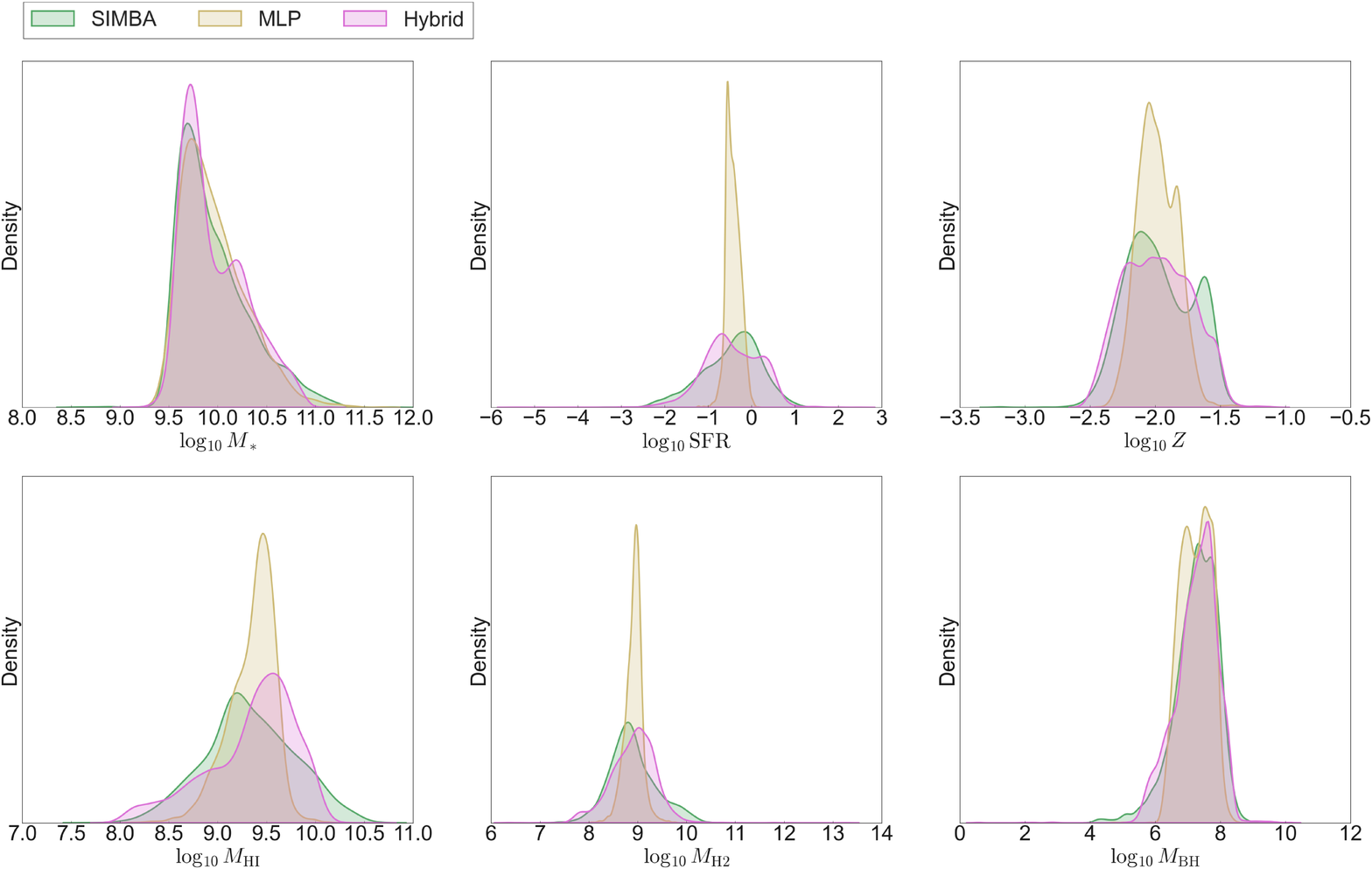}
\caption{Density plots for predictions on all six baryonic target properties. The panels show separate kernel density estimates for stellar mass ($M_*$), star formation rate ($\mathrm{SFR}$), metallicity ($Z$), neutral hydrogen ($M_\mathrm{HI}$), molecular hydrogen ($M_\mathrm{H2}$), and black hole mass ($M_\mathrm{BH}$). The corresponding values from the underlying \simba test dataset are plotted in yellow, baseline multilayer perceptron predictions in green, and results for the hybrid model are in red.}
\label{fig:figure_7}
\end{figure*}

In this section, we further analyze the quality of model outputs. Similar to our visualizations in Section~\ref{sec:predictions}, we wish to compare the model predictions to the underlying target data from the \simba suite of cosmological simulations, but separately for different baryonic properties. Fig~\ref{fig:figure_7} shows kernel density estimates for all six target properties. For the bandwidth optimization, we make use of Scott's rule as the default heuristic in a variety of statistical software packages for a dataset length $|X|$ with a given dimensionality,
\begin{eqnarray}
\hat{\beta} = |X|^{- (\dim(X) + 4)^{-1}}.
\label{eq:scott}
\end{eqnarray}

In addition, we also plot the same for the baseline multilayer perception used in previous comparisons. These density plots reveal a reasonably close agreement regarding scatter between \simba and our model's predictions for all baryonic properties. While the baseline model is already performing well on the stellar mass and, to a degree, on the black hole mass, it visibly struggles with the scatter of the remainder of the target properties.

Our hybrid model shows major improvements in the distributions of SFR, $Z$, $M_\mathrm{HI}$, and $M_\mathrm{H2}$. For the multilayer perceptron, the predicted distributions form a sharp peak around the mean, demonstrating especially difficult-to-retrieve scatter for the star formation rate and the molecular hydrogen mass. While the injected physical knowledge on the SHMR does not aid with scatter predictions but instead with the accuracy of associated properties, the KLD loss component contains extra information on distributions that allows for the production of the necessary scatter.

To quantify these improvements, we compare the baseline model and our hybrid approach in Table~\ref{tab:table_3}, providing MSE, $R^2$, $\rho$ and KLD values as for previous comparisons. Here, we can see the expectedly high correlations for $M_*$, $M_\mathrm{BH}$, and, to a lesser degree, $Z$, comparable to prior research on different hybrid approaches in this area \citep{Moews2021}. 

\begin{table*}
    \centering
    \caption{Statistical validation for the full experimental run of hybrid model. The table lists the MSE, the coefficient of determination ($R^2$), Pearson's correlation coefficient ($\rho$), and the KLD for each output variable.}
    \begin{tabular}{l c c c c c c c c c}
    \hline
    \multicolumn{1}{c}{} & \multicolumn{4}{c}{MLP} & \multicolumn{1}{c}{} & \multicolumn{4}{c}{Hybrid} \\
    \cline{2-5} \cline{7-10}
    Variable & MSE & $R^2$ & $\rho$ & KLD & & MSE & $R^2$ & $\rho$ & KLD \\
    \hline
    $M_*$ & 0.023 & 0.827 & 0.901 & 0.049 & & 0.020 & 0.847 & 0.920 & 0.045 \\
    $\mathrm{SFR}$ & 0.638 & 0.050 & 0.267 & 0.245 & & 0.423 & 0.080 & 0.284 & 0.226\\
    $Z$ & 0.052 & 0.277 & 0.527 & 0.114 & & 0.044 & 0.352 & 0.593 & 0.078\\
    $M_\mathrm{HI}$ & 0.221 & 0.215 & 0.450 & 0.215 & & 0.162 & 0.235 & 0.480 & 0.191\\
    $M_\mathrm{H2}$ & 0.308 & 0.087 & 0.284 & 0.293 & & 0.208 & 0.097 & 0.313 & 0.290\\
    $M_\mathrm{BH}$ & 0.276 & 0.466 & 0.682 & 0.090 & & 0.214 & 0.502 & 0.709 & 0.091\\
    \hline
    \end{tabular}
    \label{tab:table_3}
\end{table*}

While the latter does not use neural network architectures or direct injection of information into a loss function, they build an additional analytic model into the prediction pipeline and use full largest-progenitor merger trees, which demonstrates the capability of our loss function extensions.

Despite the correct prediction of mean relations, results for SFR, $M_\mathrm{HI}$, and $M_\mathrm{H2}$ are not quite as perfect when viewed next to the remaining baryonic properties. One reason is that we use a single neural network model to predict all variables simultaneously instead of using separate models. The downside is that the model cannot focus solely on a one-dimensional prediction, but this is not desirable, as it also bars the model from learning the connections between different target variables. The result, however, is that properties with a smaller correlation coefficient in terms of input variables, as shown in Figure~\ref{fig:figure_4}, are more difficult to recover.

For the star formation rate and both hydrogen masses, we can see in Figure~\ref{fig:figure_6} that there is more scatter at higher dark matter halo masses for these targets in particular, which usually have higher scatter compared to $M_*$, Z, and $M_\mathrm{BH}$ \citep{Shankar2017}. Another reason for prediction errors is likely found in the normal assumption. To calculate the KLD loss and implant it into the backpropagation, we assume a 6D Gaussian distribution, which is not completely true for our data. More complex approximations, however, are beyond the scope of this initial study due to often prohibitive computational costs and the challenge of maintaining differentiability but are discussed in Section~\ref{sec:discussion}.

There are potential solutions, which include the injection of more information, for example, on scatter relations, into the loss function. Another pathway is applying more data, such as merger tree information, into the input features. The main purpose of this paper is to show that PINNs, which make use of extra physical knowledge, together with a second distributional loss component, can improve model performance in this challenging area and demonstrate the utility that current developments in deep learning approaches can provide for the simulation of baryonic properties.  

\section{Discussion and limitations}
\label{sec:discussion}

The application of modern machine learning methods to the completion of $N$-body information has emerged as a growing area of interest in recent years. Here, the sometimes mentioned unreasonable effectiveness of tree-based ensembles, most commonly random forests and extra trees, is that these models are comparatively simple in their functionality, and yet, as covered in Section~\ref{sec:introduction}, are frequently found to outperform neural network architectures in this area. For standard feed-forward frameworks, the universal approximation theorem even guarantees the ability to represent arbitrary functions under very limited assumptions \citep{Cybenko1989, Hornik1989, Maiorov1999}. These theoretical capabilities do not, however, touch upon the `learnability' of these functions, as there are a wide variety of hyperparameters to consider.

While tree-based ensembles are simple, they too are not immune to such limitations, and more recent research has discussed expected error rates as non-monotonous functions concerning the number of ensemble constituents to be set by the user \citep{Probst2018}. With the first application of machine learning to this specific area having arguably started with \citet{Kamdar2016}, recent works such as \citet{Moster2021}, \citet{Jespersen2022}, and \citet{Stiskalek2022} have demonstrated the performance potential of suitably chosen neural network architectures.

One issue that remains is the limitations set by the data used for training and prediction in supervised machine learning models. A popular adage from the early days of computer science still widely used is `garbage in, garbage out', expressing the constraints on arbitrarily powerful models by insufficient datasets. While the results of cosmological simulations are, of course, by no means garbage, they are also limited by the physics they implement and the chosen assumptions and simplifications. PINNs provide a way to go beyond these ingredients, forcing the model to consider further explicitly specified physical relationships. 

Extensions of this exploratory study on the application of PINNs in this area could address complications from the black hole seeding process, including zero-mass black hole values in the dataset. As this inclusion is likely to distort the predictive distributions due to a sharp cut-off, the question is whether models can be trained to sufficiently differentiate between these two regimes or an additional distinction step is required. A secondary discriminative model is a potential solution for this challenge, which would take the place of a filter that assigns feature vectors to two different models; one for zero-mass black hole predictions and one for the remaining majority of data points. While beyond the scope of this study, this could avoid smearing effects in the space between those two cases.

An additional extension to the presented research is the inclusion of other hydrodynamic simulations, as varying implementation details can lead to differences in the produces datasets. One example for this is the presence of quenching in \simba, which impacts the correlation between SFR and $M_*$. Depending on the goal of work in this area, for example when focusing on the main sequence, other simulations can be considered to suit these needs.

Our additional implementation of distributional compliance as part of the loss function puts a further constraint on the model, which targets scatter fidelity. The direct incorporation of this additional knowledge into the architecture also goes beyond prior work on the inclusions of analytic models, which relies on physical computations outside of and before the application of machine learning models, as previously implemented by \citet{Moews2021}.

At the same time, the predictive power of our architecture relies on the suitability of the included loss function components in terms of both the physical domain knowledge provided and the choice of a density approximation method. As covered in Section~\ref{sec:shmr_fitting}, the inclusion of baryon conversion efficiency through the relationship between $M_*$ and $M_h$ presents a simplification. This is especially the case due to stellar feedback resulting in lowered efficiency at lower masses and AGN feedback resulting in lowered efficiency at high masses. For future research, if this particular constraint is to be kept as part of the model, this should be further refined to more closely follow physical theory and represents a limitation.

The natural extension of our work, although beyond the scope of this paper, is the identification and analysis of additional domain knowledge to be injected into the loss function, thus allowing the model to root its learning in a more complete set of physics. Another pathway is replacing our normal assumption with more complex approximations that allow for non-Gaussian features and multimodal distributions.

\begin{table}  
    \centering
    \caption{Dataset size and input variables for related works and this paper. Here, $S_{x,y,z}$ denotes the different components of the spin, $V_\mathrm{c}$ the maximum circular velocity in the subhalo, $N_{h}$ the number of dark matter particles bound to the subhalo, $\lambda_h$ the halo spin, and $\rho_h$ the local halo density. Sets for additional information at higher redshifts are indicated using braces.}
    \begin{tabular}{l  r c}
    \hline
       & size  & input  \\
    \hline
    \citet{Kamdar2016} & $249370$ & $M_{h}$, $S_x$, $S_y$, $S_z$, $\sigma_h$, $N_{h}$, $V_\mathrm{c}$ \\
    \citet{Agarwal2018} & $3400$ & $\{M_h\}_i$, $\{\rho_h\}_j$, $\lambda_h$, $\sigma_h$ \\
    \citet{Moews2021} & $13132$ & $\{M_h\}_i$, $r_h$, $\sigma_h$\\
    Dai et al. (this work)  & $14247$ & $M_h$, $r_h$, $\sigma_h$\\
    \hline
    \end{tabular}
    \label{tab:table_4}
\end{table}

In Table~\ref{tab:table_4}, we list the dataset sizes and different dark matter properties used as inputs in similar works providing the same performance metrics \citep{Kamdar2016, Agarwal2018, Moews2021}. In contrast to this paper, all three comparable studies employ tree-based models, with the largest dataset used by \citet{Kamdar2016}. The latter also provide their model, in addition to $M_h$ and $\sigma_h$, with the three spin components, number of dark matter particles bound to the subhalo, and maximum circular velocity in the subhalo, but omit $r_h$ as used in our work. This supplies additional information on the dark matter halo of their model.

Similarly, \citet{Agarwal2018} do not include $r_h$, but add the halo local density, with $\{M_h\}_i\}$ denoting not only the current halo mass at $z = 0$ but also the five preceding snapshots at higher redshifts, which provides information on the recent merger history. Conversely, $\{\rho\}_j$ indicates the set of nearby halo mass densities within radii $r \in \{200, 500, 1000\}$ (in kpc) focused on the halo's mass center, thus providing additional information in a similar way. That being said, their dataset is the smallest in this line-up. 

The work most closely aligned with ours in terms of inputs is \citet{Moews2021}, although largest-progenitor merger trees over the entirety of a halo's evolution are provided instead of only the current halo mass. More importantly, their hybrid approach uses an analytic formalism to pre-compute a subset of baryonic properties with a physical model, feeding the merger trees both into the latter and the subsequent machine learning model.

\begin{table}
    \centering
    \caption{Performance metrics for related works. The table lists the coefficient of determination ($R^2$) and Pearson's correlation coefficient ($\rho$) for each output variable. Here, in addition to the variables described throughout this paper, $M_\mathrm{gas}$ denotes the total gas mass of a subhalo in the indicated paper.}
    \begin{tabular}{l c c c c c c c c}
    \hline
    \multicolumn{1}{c}{} & \multicolumn{2}{c}{Kamdar et al.} & \multicolumn{1}{c}{} & \multicolumn{2}{c}{Agarwal et al.} & \multicolumn{1}{c}{} & \multicolumn{2}{c}{Moews et al.}\\
    \cline{2-3} \cline{5-6} \cline{8-9}
    Variable & $R^2$ & $\rho$ & & $R^2$ & $\rho$ & & $R^2$ & $\rho$ \\
    \hline
    $M_*$ & 0.91 & 0.95 & & 0.90 & 0.95 & & 0.82 & 0.94 \\
    $\mathrm{SFR}$ & 0.63 & 0.79 & & 0.55 & 0.74 & & 0.73 & 0.87 \\
    $Z$ & 0.93 & 0.96 & & 0.73 & 0.86 & & 0.21 & 0.66 \\
    $M_\mathrm{gas}$ & 0.67 & 0.85 &  & - & - &  & - &- \\
    $M_\mathrm{HI}$ & - & - & & 0.35  & 0.59 & & 0.36 & 0.65 \\
    $M_\mathrm{H2}$ & - & - & & 0.51 & 0.71 & & 0.54 & 0.75\\
    $M_\mathrm{BH}$ & 0.72 & 0.85 & & - & - & & 0.71 & 0.88\\
    \hline
    \end{tabular}
    \label{tab:table_5}
\end{table}

In Table~\ref{tab:table_5}, we list the performance metrics, $R^2$ and $\rho$, for these related works to enable a discussion of differences in data and model approach. With regard to \citet{Kamdar2016}, our model yields a close performance on $M_{*}$ and $M_\mathrm{BH}$, but with a less robust prediction of SFR and $Z$. This could be due to the size of the \illustris\ dataset used by the authors, which is about eight times larger than our available \simba\ data, as more suitable data usually leads to a better performance in machine learning models \citep{Zhou2014}. 

\citet{Agarwal2018} use \mufasa, the predecessor of \simba, and benefits from their inclusion of prior halo masses and halo mass densities at different distance radii. Our model's results are comparable for $M_*$ and $M_\mathrm{HI}$, with a slight underperformance for $Z$, but there is a notable degradation in $M_\mathrm{H2}$ and SFR predictions. Based on the correlation matrix in Figure~\ref{fig:figure_4}, these two properties are highly correlated, meaning that an improvement in one in future research should impact the other. One potential reason for this degradation is the larger scatter in \simba\ higher halo masses, while \citet{Agarwal2018} pre-select their galaxies to be star-forming, which strongly reduces the scatter at high masses.

Lastly, \citet{Moews2021} develop a hybrid approach that combines an extra trees ensemble with the equilibrium model and incorporates merger trees into the latter. The same data and inputs are used in both works, save for the largest-progenitor merger trees that are fed into both the physical and machine learning models. While our model achieves close results on $M_*$, Z, $M_\mathrm{HI}$, and, to a lesser degree, $M_\mathrm{BH}$, metrics of SFR and $M_{H2}$ are lower. Here, the difficulty in predicting the star formation rate due to the large scatter could be data-driven, while \citet{Moews2021} utilize a physical model for this property. At the same time, our model outperforms these results on $Z$, which reflects the equilibrium model's difficulty with this quantity.

Overall, unlike our method focusing on the retrieval of accurate scatter, the above models are optimized for accuracy and make use of more information on the dark matter halo. The results of both \citet{Agarwal2018} and \citet{Moews2021} show a considerably less accurate scatter retrieval than can be seen in our experiments, with noticeable underprediction of the tails. The aim of this work is not to replace existing research but to demonstrate modifications that can increase the predictive ability of deep learning models in this area. Since the dataset, model, and purpose are different, this comparison provides a direction for future research targeting the combination of these strengths. With regard to the model architecture, the closer approximation of scatter along each variable's range instead of a general distribution parameter per variable could further improve the retrieval of scatter. While beyond the scope of this work, this should be combined with a closer look at alternative activation functions to potentially alleviate the wave-like nature of some predictions as described in Section~\ref{sec:predictions}.

For follow-up research not using PINNs, or machine learning models that do not require the loss function to be differentiable for backpropagation as described in Section~\ref{sec:kld_loss}, we recommend extensions to the comparison of target and prediction distributions. This work uses the KLD to calculate the difference between those distributions under the normal assumption. While a reasonably close proposition for the data used in our experiments, this limits our model's effectiveness when directly transferred to markedly non-Gaussian datasets. In such cases, the spread of predictions through the variance would still be enforced, but in the case of, for example, a starkly multimodal distribution, the recreation of these distributional features would not be a major component of the optimization. This is also the reason for concerns covered in Section~\ref{sec:kld_loss} with regard to potential negative effects on the mean accuracy, for example in case of strongly skewed distributions. Such negative effects are not present in our results, but could be argued to be a more relevant concern for starkly non-normal distributions in other research.

One point that should be discussed beforehand is the improvement in MSE, $R^2$ and $\rho$ values in Table~\ref{tab:table_1} in Section~\ref{sec:shmr_fitting}. While no decrease in these key performance indices, as covered above, shows that the distributional constraint allows for retaining the level of accuracy achieved only through the inclusion of the SHMR, an improvement is not immediately intuitive, as the KLD component targets the scatter reproduction, not the predictive accuracy. The underprediction of tails is a common problem in the application of machine learning models to predict baryonic properties from dark matter halo information; this is not only the case in this research, but also in other publications, as mentioned above.

One possibility is that a more accurate distribution that places the data and predictions within the same range reduces, on average, the point-by-point error through a more accurate distribution width. Another reason could be found in the reduced underprediction of tails, which can be viewed as a subcase of the first possibility, as it results in a decreased potential for errors stemming from a lack of accurate predictions in these regions.

As the comparison needs to be reasonably fast, approaches such as Bayesian mixture models as well as associated methods that are more complex \citep[see, for example,][]{Moews2020} are likely to slow the training process down too much. As a compromise, a kernel density estimate (KDE), also known as the Parzen-Rosenblatt window after \citet{Rosenblatt1956} and \citet{Parzen1962}, can be used, although this is limited to lower dimensionalities. We propose two ways to circumvent the latter limitation. The first is applying one-dimensional KDEs on a variable-by-variable basis and then averaging the Kullback-Leibler divergences between these estimates. While the advantage is the good fit in $\mathbb{R}^1$, this can lead to a subset of variables not being forced to follow the target distribution as long as the KLD average remains small.

The alternative is to make use of dimensionality reduction methods such as principal component analysis, which collapses the $n$-dimensional coordinate space, for $n$ variables, into orthogonal vectors ranked by their ability to explain the variance \citep[for a recent overview, see][]{Joliffe2016}. Reducing the coordinate space to a subspace in $\mathbb{R}^2$ would, for example, still allow for reasonably good KDE approximations while retaining each variable's contribution to a combined KLD. The same line of thought does, of course, also apply to other density approximation and dimensionality reduction methods.

\section{Conclusion}
\label{sec:conclusion}

In this paper, we transfer the paradigm of physics-informed neural networks to predicting baryonic properties for associated dark matter halo variables. We adapt this approach in two different ways. The first includes the stellar-to-halo mass relation, a double power law previously used to parameterize the instantaneous baryon conversion efficiency. While this is a more established way to include physical theory into PINNs, our second extension is the enforcement of baryonic scatter in simulations under a normal assumption using the Kullback-Leibler divergence between the underlying cosmological simulation and predictions. In doing so, and separately from improvements in the mean accuracy of predictions, we solve the common problem of scatter reproduction in this area, which is merged directly into the machine learning model.

We first test the improvement for the more traditional approach to PINNs, meaning the injection of physical domain knowledge into the loss function, and demonstrate a positive effect on the model's performance. The hybrid approach combines these strengths by including the measurement of distributional differences and outperforms the standard PINN model. Subsequent tests of the scatter retrieval show more faithful reproductions for baryonic properties. These improvements in scatter are especially notable for molecular hydrogen masses and star formation rates but can also be seen for neutral hydrogen masses and metallicities. In particular, our model successfully recovers the fundamental metallicity relation.

Our experiments demonstrate that PINNs, a rapidly expanding area of research across various subfields of physics, offer a way to directly bake theoretical constraints and distributional adherence into neural network architectures when painting baryonic properties into galactic dark matter halos. As such, they can be used to complete cosmological $N$-body simulations based on full hydrodynamic simulation suites, although this comes with the same caveats as other research in this area. 

The inference of physics from simulations operates under the assumption that such simulations are a sufficiently close approximation of the real world. Any machine learning models learning from those simulations are subject to the same assumption. That said, including physical models in the learning process enables these algorithms to include domain information beyond the underlying cosmological simulations.

Potential follow-ups include additional physical models specific to galaxy formation and evolution into the loss function, and further constraints based on observational data or other simulations to diversify the data sources. Our presented framework is widely applicable to large-scale cosmological simulations and the study of the utility and effect of physical domain knowledge on galaxy evolution emulators. It provides a further piece in the puzzle of fully using modern machine learning in astrophysics.

\section*{Acknowledgments}
This work received support from the Hewlett Packard Enterprise Data Science Institute at the University of Houston. BM acknowledges support from a McWilliams Fellowship at Carnegie Mellon University. \simba\ was run on the DiRAC@Durham facility managed by the Institute for Computational Cosmology on behalf of the STFC DiRAC HPC Facility. The equipment was funded by BEIS (Department for Business, Energy \& Industrial Strategy) capital funding via STFC capital grants ST/P002293/1, ST/R002371/1 and ST/S002502/1, Durham University, and STFC operations grant ST/R000832/1. DiRAC is part of the National e-Infrastructure. 

\section*{Data and Software Availability}

The \simba simulation data and galaxy catalogs underlying this article are publicly available at \url{https://simba.roe.ac.uk}.

\bibliographystyle{mnras}
\bibliography{ref}

\begin{thebibliography}{}
\makeatletter
\relax
\def\mn@urlcharsother{\let\do\@makeother \do\$\do\&\do\#\do\^\do\_\do\%\do\~}
\def\mn@doi{\begingroup\mn@urlcharsother \@ifnextchar [ {\mn@doi@}
  {\mn@doi@[]}}
\def\mn@doi@[#1]#2{\def\@tempa{#1}\ifx\@tempa\@empty \href
  {http://dx.doi.org/#2} {doi:#2}\else \href {http://dx.doi.org/#2} {#1}\fi
  \endgroup}
\def\mn@eprint#1#2{\mn@eprint@#1:#2::\@nil}
\def\mn@eprint@arXiv#1{\href {http://arxiv.org/abs/#1} {{\tt arXiv:#1}}}
\def\mn@eprint@dblp#1{\href {http://dblp.uni-trier.de/rec/bibtex/#1.xml}
  {dblp:#1}}
\def\mn@eprint@#1:#2:#3:#4\@nil{\def\@tempa {#1}\def\@tempb {#2}\def\@tempc
  {#3}\ifx \@tempc \@empty \let \@tempc \@tempb \let \@tempb \@tempa \fi \ifx
  \@tempb \@empty \def\@tempb {arXiv}\fi \@ifundefined
  {mn@eprint@\@tempb}{\@tempb:\@tempc}{\expandafter \expandafter \csname
  mn@eprint@\@tempb\endcsname \expandafter{\@tempc}}}

\bibitem[\protect\citeauthoryear{Agarwal, Dav{\'e}  \& Bassett}{Agarwal
  et~al.}{2018}]{Agarwal2018}
Agarwal S.,  Dav{\'e} R.,   Bassett B.~A.,  2018, \mn@doi [\mnras]
  {10.1093/mnras/sty1169}, 478, 3410

\bibitem[\protect\citeauthoryear{Akiba, Sano, Yanase, Ohta  \& Koyama}{Akiba
  et~al.}{2019}]{optuna_2019}
Akiba T.,  Sano S.,  Yanase T.,  Ohta T.,   Koyama M.,  2019, in Proceedings of
  the 25th {ACM} {SIGKDD} International Conference on Knowledge Discovery and
  Data Mining.

\bibitem[\protect\citeauthoryear{Baugh et~al.,}{Baugh et~al.}{2018}]{Baugh2018}
Baugh C.~M.,  et~al., 2018, \mn@doi [\mnras] {10.1093/mnras/sty3427}, 483, 4922

\bibitem[\protect\citeauthoryear{{Behroozi}, {Wechsler}, {Hearin}  \&
  {Conroy}}{{Behroozi} et~al.}{2019}]{Behroozi2019}
{Behroozi} P.,  {Wechsler} R.~H.,  {Hearin} A.~P.,   {Conroy} C.,  2019,
  \mn@doi [\mnras] {10.1093/mnras/stz1182}, 488, 3143

\bibitem[\protect\citeauthoryear{{Ben-David}, Liu  \& Jackson}{{Ben-David}
  et~al.}{2015}]{Ben-David2015}
{Ben-David} A.,  Liu H.,   Jackson A.~D.,  2015, \mn@doi [\jcap]
  {10.1088/1475-7516/2015/06/051}, 2015, 051

\bibitem[\protect\citeauthoryear{Berlind \& Weinberg}{Berlind \&
  Weinberg}{2002}]{Berlind2002}
Berlind A.~A.,  Weinberg D.~H.,  2002, \mn@doi [\apj] {10.1086/341469}, 575,
  587

\bibitem[\protect\citeauthoryear{Blumenthal, Faber, Primack  \&
  Rees}{Blumenthal et~al.}{1984}]{Blumenthal1984}
Blumenthal G.,  Faber S.~M.,  Primack J.~R.,   Rees M.~J.,  1984, \mn@doi
  [\nat] {10.1038/311517a0}, 311, 517

\bibitem[\protect\citeauthoryear{Bouch{\'e} et~al.,}{Bouch{\'e}
  et~al.}{2010}]{Bouche2010}
Bouch{\'e} N.,  et~al., 2010, \mn@doi [\apj] {10.1088/0004-637X/718/2/1001},
  718, 1001

\bibitem[\protect\citeauthoryear{{Boylan-Kolchin}, Springel, White, Jenkins  \&
  Lemson}{{Boylan-Kolchin} et~al.}{2009}]{Boylan-Kolchin2009}
{Boylan-Kolchin} M.,  Springel V.,  White S. D.~M.,  Jenkins A.,   Lemson G.,
  2009, \mn@doi [\mnras] {10.1111/j.1365-2966.2009.15191.x}, 398, 1150

\bibitem[\protect\citeauthoryear{Branca \& Pallottini}{Branca \&
  Pallottini}{2022}]{Branca2022}
Branca L.,  Pallottini A.,  2022, \mn@doi [\mnras] {10.1093/mnras/stac3512},
  518, 5718

\bibitem[\protect\citeauthoryear{Campbell, van~den Bosch, Padmanabhan, Mao,
  Zentner, Lange, Jiang  \& Villarreal}{Campbell et~al.}{2018}]{Campbell2018}
Campbell D.,  van~den Bosch F.~C.,  Padmanabhan N.,  Mao Y.-Y.,  Zentner A.~R.,
   Lange J.~U.,  Jiang F.,   Villarreal A.~S.,  2018, \mn@doi [Monthly Notices
  of the Royal Astronomical Society] {10.1093/mnras/sty495}, 477, 359

\bibitem[\protect\citeauthoryear{Cao, Tinker, Mao  \& Wechsler}{Cao
  et~al.}{2020}]{Cao2020}
Cao J.,  Tinker J.~L.,  Mao Y.,   Wechsler R.~H.,  2020, \mn@doi [\mnras]
  {10.1093/mnras/staa2644}, 498, 5080

\bibitem[\protect\citeauthoryear{Carleo, Cirac, Cranmer, Daudet, Schuld,
  Tishby, {Vogt-Maranto}  \& Zdeborov{\'a}}{Carleo et~al.}{2019}]{Carleo2019}
Carleo G.,  Cirac I.,  Cranmer K.,  Daudet L.,  Schuld M.,  Tishby N.,
  {Vogt-Maranto} L.,   Zdeborov{\'a} L.,  2019, \mn@doi [Rev. Mod. Phys.]
  {10.1103/RevModPhys.91.045002}, 91, 045002

\bibitem[\protect\citeauthoryear{Cattaneo et~al.,}{Cattaneo
  et~al.}{2017}]{Cattaneo2017}
Cattaneo A.,  et~al., 2017, \mn@doi [\mnras] {10.1093/mnras/stx1597}, 471, 1401

\bibitem[\protect\citeauthoryear{Cole, Lacey, Baugh  \& Frenk}{Cole
  et~al.}{2000}]{Cole2000}
Cole S.,  Lacey C.~G.,  Baugh C.~M.,   Frenk C.~S.,  2000, \mn@doi [\mnras]
  {10.1046/j.1365-8711.2000.03879.x}, 319, 168

\bibitem[\protect\citeauthoryear{Cornell, Ncube  \& Harmsen}{Cornell
  et~al.}{2022}]{Cornell2022}
Cornell A.~S.,  Ncube A.,   Harmsen G.,  2022, \mn@doi [\prd]
  {10.1103/PhysRevD.106.124047}, 106, 124047

\bibitem[\protect\citeauthoryear{Croton et~al.,}{Croton
  et~al.}{2016}]{Croton2016}
Croton D.~J.,  et~al., 2016, \mn@doi [\apjs] {10.3847/0067-0049/222/2/22}, 222,
  22

\bibitem[\protect\citeauthoryear{Cui et~al.,}{Cui et~al.}{2018}]{Cui2018}
Cui W.,  et~al., 2018, \mn@doi [\mnras] {10.1093/mnras/sty2111}, 480, 2898

\bibitem[\protect\citeauthoryear{Cuomo, Di~Cola, Giampaolo, Rozza, Raissi  \&
  Piccialli}{Cuomo et~al.}{2022}]{Cuomo2022}
Cuomo S.,  Di~Cola V.,  Giampaolo F.,  Rozza G.,  Raissi M.,   Piccialli F.,
  2022, \mn@doi [J. Sci. Comput.] {10.1007/s10915-022-01939-z}, 92, 88

\bibitem[\protect\citeauthoryear{Cybenko}{Cybenko}{1989}]{Cybenko1989}
Cybenko G.,  1989, \mn@doi [Math. Control Signals Syst.] {10.1007/BF02551274},
  2, 303

\bibitem[\protect\citeauthoryear{Dav{\'e}, Finlator  \& Oppenheimer}{Dav{\'e}
  et~al.}{2012}]{Dave2012}
Dav{\'e} R.,  Finlator K.,   Oppenheimer B.~D.,  2012, \mn@doi [\mnras]
  {10.1111/j.1365-2966.2011.20148.x}, 421, 98

\bibitem[\protect\citeauthoryear{Dav{\'e}, Thompson  \& Hopkins}{Dav{\'e}
  et~al.}{2016}]{Dave2016}
Dav{\'e} R.,  Thompson R.,   Hopkins P.~F.,  2016, \mn@doi [\mnras]
  {10.1093/mnras/stw1862}, 462, 3265

\bibitem[\protect\citeauthoryear{Dav{\'e}, {Angl{\'e}s-Alc{\'a}zar}, Narayanan,
  Li, Rafieferantsoa  \& Appleby}{Dav{\'e} et~al.}{2019}]{Dave2019}
Dav{\'e} R.,  {Angl{\'e}s-Alc{\'a}zar} D.,  Narayanan D.,  Li Q.,
  Rafieferantsoa M.~H.,   Appleby S.,  2019, \mn@doi [\mnras]
  {10.1093/mnras/stz937}, 486, 2827

\bibitem[\protect\citeauthoryear{Deiana, Tran, Agar  \& et. al.}{Deiana
  et~al.}{2022}]{Deiana2022}
Deiana A.,  Tran N.,  Agar J.,   et. al. 2022, \mn@doi [Front. Big Data]
  {10.3389/fdata.2022.787421}, 5, 787421

\bibitem[\protect\citeauthoryear{Desmond, Mao, Wechsler, Crain  \&
  Schaye}{Desmond et~al.}{2017}]{Desmond2017}
Desmond H.,  Mao Y.,  Wechsler R.~H.,  Crain R.~A.,   Schaye J.,  2017, \mn@doi
  [MNRAS Let.] {10.1093/mnrasl/slx093}, 471, L11

\bibitem[\protect\citeauthoryear{Dissanayake \& Phan-Thien}{Dissanayake \&
  Phan-Thien}{1994}]{Dissanayake1994}
Dissanayake M. W. M.~G.,  Phan-Thien N.,  1994, \mn@doi [Commun. Numer. Methods
  Eng.] {doi.org/10.1002/cnm.1640100303}, 10, 195

\bibitem[\protect\citeauthoryear{Dolag, Borgani, Schindler, Diaferio  \&
  Bykov}{Dolag et~al.}{2008}]{Dolag2008}
Dolag K.,  Borgani S.,  Schindler S.,  Diaferio A.,   Bykov A.~M.,  2008,
  \mn@doi [\ssr] {10.1007/s11214-008-9316-5}, 134, 229

\bibitem[\protect\citeauthoryear{Dubois, Peirani, Pichon, Devriendt, Gavazzi,
  Welker  \& Volonteri}{Dubois et~al.}{2016}]{Dubois2016}
Dubois Y.,  Peirani S.,  Pichon C.,  Devriendt J.,  Gavazzi R.,  Welker C.,
  Volonteri M.,  2016, \mn@doi [\mnras] {10.1093/mnras/stw2265}, 463, 3948

\bibitem[\protect\citeauthoryear{Efstathiou, Davis, White  \& Frenk}{Efstathiou
  et~al.}{1985}]{Efstathiou1985}
Efstathiou G.,  Davis M.,  White S. D.~M.,   Frenk C.~S.,  1985, \mn@doi
  [\apjs] {10.1086/191003}, 57, 241

\bibitem[\protect\citeauthoryear{Ferdosi, Buddelmeijer, Trager, Wilkinson  \&
  Roerdink}{Ferdosi et~al.}{2011}]{Ferdosi2011}
Ferdosi B.~J.,  Buddelmeijer H.,  Trager S.~C.,  Wilkinson M. H.~F.,   Roerdink
  J. B. T.~M.,  2011, \mn@doi [\aap] {10.1051/0004-6361/201116878}, 531, A114

\bibitem[\protect\citeauthoryear{Frenk \& White}{Frenk \&
  White}{2012}]{Frenk2012}
Frenk C.~S.,  White S. D.~M.,  2012, \mn@doi [Ann. Phys.]
  {10.1002/andp.201200212}, 524, 507

\bibitem[\protect\citeauthoryear{Fussell \& Moews}{Fussell \&
  Moews}{2019}]{Fussell2019}
Fussell L.,  Moews B.,  2019, \mn@doi [\mnras] {10.1093/mnras/stz602}, 485,
  3203

\bibitem[\protect\citeauthoryear{Genel et~al.,}{Genel et~al.}{2014}]{Genel2014}
Genel S.,  et~al., 2014, \mn@doi [\mnras] {10.1093/mnras/stu1654}, 445, 175

\bibitem[\protect\citeauthoryear{Hastie, Tibshirani  \& Friedman}{Hastie
  et~al.}{2009}]{Hastie2009}
Hastie T.,  Tibshirani R.,   Friedman J.,  2009, {The elements of statistical
  learning: data mining, inference and prediction}, 2 edn.
New York, USA: Springer

\bibitem[\protect\citeauthoryear{Hatton, Devriendt, Ninin, Bouchet, Guiderdoni
  \& Vibert}{Hatton et~al.}{2003}]{Hatton2003}
Hatton S.,  Devriendt J. E.~G.,  Ninin S.,  Bouchet F.~R.,  Guiderdoni B.,
  Vibert D.,  2003, \mn@doi [\mnras] {10.1046/j.1365-8711.2003.05589.x}, 343,
  75

\bibitem[\protect\citeauthoryear{Hearin, Korytov, Kovacs, Benson, Aung,
  Bradshaw, Campbell  \& Collaboration)}{Hearin et~al.}{2020}]{Hearin2020}
Hearin A.,  Korytov D.,  Kovacs E.,  Benson A.,  Aung H.,  Bradshaw C.,
  Campbell D.,   Collaboration) T. L. D. E.~S.,  2020, \mn@doi [\mnras]
  {10.1093/mnras/staa1495}, 495, 5040

\bibitem[\protect\citeauthoryear{Hee, Vázquez, Handley, Hobson  \&
  Lasenby}{Hee et~al.}{2016}]{Hee2016}
Hee S.,  Vázquez J.~A.,  Handley W.~J.,  Hobson M.~P.,   Lasenby A.~N.,  2016,
  \mn@doi [\mnras] {10.1093/mnras/stw3102}, 466, 369

\bibitem[\protect\citeauthoryear{{Hopkins}}{{Hopkins}}{2015}]{Hopkins2015}
{Hopkins} P.~F.,  2015, \mn@doi [\mnras] {10.1093/mnras/stv195}, 450, 53

\bibitem[\protect\citeauthoryear{Hornik, Stinchcombe  \& White}{Hornik
  et~al.}{1989}]{Hornik1989}
Hornik K.,  Stinchcombe M.,   White H.,  1989, \mn@doi [Neural Netw.]
  {10.1016/0893-6080(89)90020-8}, 2, 359

\bibitem[\protect\citeauthoryear{Jespersen, Cranmer, Melchior, Ho, Somerville
  \& Gabrielpillai}{Jespersen et~al.}{2022}]{Jespersen2022}
Jespersen C.~K.,  Cranmer M.,  Melchior P.,  Ho S.,  Somerville R.~S.,
  Gabrielpillai A.,  2022, \mn@doi [\apj] {10.3847/1538-4357/ac9b18}, 941, 7

\bibitem[\protect\citeauthoryear{Jin, Zhang, Zhu, Tang  \& Karniadakis}{Jin
  et~al.}{2020}]{Jin2020}
Jin P.,  Zhang Z.,  Zhu A.,  Tang Y.,   Karniadakis G.~E.,  2020, \mn@doi
  [Neural Netw.] {https://doi.org/10.1016/j.neunet.2020.08.017}, 132, 166

\bibitem[\protect\citeauthoryear{Jo \& Kim}{Jo \& Kim}{2019}]{Jo2019}
Jo Y.,  Kim J.,  2019, \mn@doi [\mnras] {10.1093/mnras/stz2304}, 489, 3565

\bibitem[\protect\citeauthoryear{Jolliffe \& Cadima}{Jolliffe \&
  Cadima}{2016}]{Joliffe2016}
Jolliffe I.~T.,  Cadima J.,  2016, \mn@doi [Philos. Trans. R. Soc. A]
  {10.1098/rsta.2015.0202}, 374, 20150202

\bibitem[\protect\citeauthoryear{Kamdar, Turk  \& Brunner}{Kamdar
  et~al.}{2016}]{Kamdar2016}
Kamdar H.~M.,  Turk M.~J.,   Brunner R.~J.,  2016, \mn@doi [\mnras]
  {10.1093/mnras/stv2981}, 457, 1162

\bibitem[\protect\citeauthoryear{Karniadakis, Kevrekidis, Lu, Perdikaris, Wang
  \& Yang}{Karniadakis et~al.}{2021}]{Karnadiakis2021}
Karniadakis G.~E.,  Kevrekidis I.~G.,  Lu L.,  Perdikaris P.,  Wang S.,   Yang
  L.,  2021, \mn@doi [Nat. Rev. Phys.] {10.1038/s42254-021-00314-5}, 3, 422–

\bibitem[\protect\citeauthoryear{Kashefi, Rempe  \& Guibas}{Kashefi
  et~al.}{2021}]{Kashefi2021}
Kashefi A.,  Rempe D.,   Guibas L.~J.,  2021, \mn@doi [Phys. Fluids]
  {10.1063/5.0033376}, 33, 027104

\bibitem[\protect\citeauthoryear{Klypin, {Trujillo-Gomez}  \& Primack}{Klypin
  et~al.}{2011}]{Klypin2011}
Klypin A.~A.,  {Trujillo-Gomez} S.,   Primack J.,  2011, \mn@doi [\apj]
  {10.1088/0004-637X/740/2/102}, 740, 102

\bibitem[\protect\citeauthoryear{Kravtsov, Vikhlinin  \&
  Meshcheryakov}{Kravtsov et~al.}{2018}]{Kravtsov2018}
Kravtsov A.~V.,  Vikhlinin A.~A.,   Meshcheryakov A.~V.,  2018, \mn@doi
  [Astron. Lett.] {10.1134/s1063773717120015}, 44, 8

\bibitem[\protect\citeauthoryear{Krumholz \& Dekel}{Krumholz \&
  Dekel}{2012}]{Krumholz2012}
Krumholz M.~R.,  Dekel A.,  2012, \mn@doi [\apj] {10.1088/0004-637X/753/1/16},
  753, 16

\bibitem[\protect\citeauthoryear{Kullback \& Leibler}{Kullback \&
  Leibler}{1951}]{Kullback1951}
Kullback S.,  Leibler R.~A.,  1951, \mn@doi [Ann. Math. Stat.]
  {10.1214/aoms/1177729694}, 22, 79

\bibitem[\protect\citeauthoryear{Lagaris, Likas  \& Fotiadis}{Lagaris
  et~al.}{1998}]{Lagaris1998}
Lagaris I.,  Likas A.,   Fotiadis D.,  1998, \mn@doi [IEEE Trans. Neural Netw.]
  {10.1109/72.712178}, 9, 987

\bibitem[\protect\citeauthoryear{LeCun \& Bengio}{LeCun \&
  Bengio}{1995}]{LeCun1995}
LeCun Y.,  Bengio Y.,  1995, in Arbib M.~A.,  ed., , Handbook of Brain Theory
  and Neural Networks.
MIT Press, p.~3361

\bibitem[\protect\citeauthoryear{Lehmann, Mao, Becker, Skillman  \&
  Wechsler}{Lehmann et~al.}{2016}]{Lehmann2016}
Lehmann B.~V.,  Mao Y.,  Becker M.~R.,  Skillman S.~W.,   Wechsler R.~H.,
  2016, \mn@doi [\apj] {10.3847/1538-4357/834/1/37}, 834, 37

\bibitem[\protect\citeauthoryear{Lovell, Wilkins, Thomas, Schaller, Baugh,
  Fabbian  \& Bah{\'e}}{Lovell et~al.}{2021}]{Lovell2021}
Lovell C.~C.,  Wilkins S.~M.,  Thomas P.~A.,  Schaller M.,  Baugh C.~M.,
  Fabbian G.,   Bah{\'e} Y.,  2021, \mn@doi [\mnras] {10.1093/mnras/stab3221},
  509, 5046

\bibitem[\protect\citeauthoryear{Lu et~al.,}{Lu et~al.}{2014}]{Lu2014}
Lu Y.,  et~al., 2014, \mn@doi [\apj] {10.1088/0004-637X/795/2/123}, 795, 123

\bibitem[\protect\citeauthoryear{Maiorov \& Pinkus}{Maiorov \&
  Pinkus}{1999}]{Maiorov1999}
Maiorov V.,  Pinkus A.,  1999, \mn@doi [Neurocomputing]
  {10.1016/S0925-2312(98)00111-8}, 25, 81

\bibitem[\protect\citeauthoryear{Martin \& Schaub}{Martin \&
  Schaub}{2022}]{Martin2022}
Martin J.,  Schaub H.,  2022, \mn@doi [Celest. Mech. Dyn. Astron.]
  {10.1007/s10569-022-10101-8}, 134, 46

\bibitem[\protect\citeauthoryear{{McGibbon} \& Khochfar}{{McGibbon} \&
  Khochfar}{2022}]{McGibbon2022}
{McGibbon} R.~J.,  Khochfar S.,  2022, \mn@doi [\mnras]
  {10.1093/mnras/stac1269}, 513, 5423

\bibitem[\protect\citeauthoryear{Mishra \& Molinaro}{Mishra \&
  Molinaro}{2021}]{Mishra2021}
Mishra S.,  Molinaro R.,  2021, \mn@doi [J. Quant. Spectrosc. Radiat. Transf.]
  {10.1016/j.jqsrt.2021.107705}, 270, 107705

\bibitem[\protect\citeauthoryear{{Mitra}, {Dav{\'e}}  \& {Finlator}}{{Mitra}
  et~al.}{2015}]{Mitra2015}
{Mitra} S.,  {Dav{\'e}} R.,   {Finlator} K.,  2015, \mn@doi [\mnras]
  {10.1093/mnras/stv1387}, 452, 1184

\bibitem[\protect\citeauthoryear{Mitra, Dav{\'e}, Simha  \& Finlator}{Mitra
  et~al.}{2017}]{Mitra2017}
Mitra S.,  Dav{\'e} R.,  Simha V.,   Finlator K.,  2017, \mn@doi [\mnras]
  {10.1093/mnras/stw2527}, 464, 2766

\bibitem[\protect\citeauthoryear{Moews \& Zuntz}{Moews \&
  Zuntz}{2020}]{Moews2020}
Moews B.,  Zuntz J.,  2020, \mn@doi [\apj] {10.3847/1538-4357/ab93cb}, 896, 98

\bibitem[\protect\citeauthoryear{Moews, de Souza, Ishida, Malz, Heneka, Vilalta
   \& Zuntz}{Moews et~al.}{2019}]{Moews2019}
Moews B.,  de Souza R.~S.,  Ishida E. E.~O.,  Malz A.~I.,  Heneka C.,  Vilalta
  R.,   Zuntz J.,  2019, \mn@doi [\prd] {10.1103/PhysRevD.99.123529}, 99,
  123529

\bibitem[\protect\citeauthoryear{Moews, Dav{\'e}, Mitra, Hassan  \& Cui}{Moews
  et~al.}{2021}]{Moews2021}
Moews B.,  Dav{\'e} R.,  Mitra S.,  Hassan S.,   Cui W.,  2021, \mn@doi
  [\mnras] {10.1093/mnras/stab1120}, 504, 4024

\bibitem[\protect\citeauthoryear{Moster, Somerville, Maulbetsch, van~den Bosch,
  Macciò, Naab  \& Oser}{Moster et~al.}{2010}]{Moster2010}
Moster B.~P.,  Somerville R.~S.,  Maulbetsch C.,  van~den Bosch F.~C.,  Macciò
  A.~V.,  Naab T.,   Oser L.,  2010, \mn@doi [\apj]
  {10.1088/0004-637X/710/2/903}, 710, 903

\bibitem[\protect\citeauthoryear{Moster, Naab  \& White}{Moster
  et~al.}{2013}]{Moster2013}
Moster B.~P.,  Naab T.,   White S. D.~M.,  2013, \mn@doi [\mnras]
  {10.1093/mnras/sts261}, 428, 3121

\bibitem[\protect\citeauthoryear{Moster, Naab  \& White}{Moster
  et~al.}{2018}]{Moster2018}
Moster B.~P.,  Naab T.,   White S. D.~M.,  2018, \mn@doi [\mnras]
  {10.1093/mnras/sty655}, 477, 1822

\bibitem[\protect\citeauthoryear{Moster, Naab, Lindstr{\"o}m  \&
  {O’Leary}}{Moster et~al.}{2021}]{Moster2021}
Moster B.~P.,  Naab T.,  Lindstr{\"o}m M.,   {O’Leary} J.~A.,  2021, \mn@doi
  [\mnras] {10.1093/mnras/stab1449}, 507, 2115

\bibitem[\protect\citeauthoryear{Nicola, Amara  \& Refregier}{Nicola
  et~al.}{2019}]{Nicola2019}
Nicola A.,  Amara A.,   Refregier A.,  2019, \mn@doi [\jcap]
  {10.1088/1475-7516/2019/01/011}, 2019, 011

\bibitem[\protect\citeauthoryear{Panter, Jimenez, Heavens  \& Charlot}{Panter
  et~al.}{2007}]{Panter2007}
Panter B.,  Jimenez R.,  Heavens A.~F.,   Charlot S.,  2007, \mn@doi [\mnras]
  {10.1111/j.1365-2966.2007.11909.x}, 378, 1550

\bibitem[\protect\citeauthoryear{Parzen}{Parzen}{1962}]{Parzen1962}
Parzen E.,  1962, \mn@doi [Ann. Math. Stat.] {10.1214/aoms/1177704472}, 33,
  1065

\bibitem[\protect\citeauthoryear{Pillepich et~al.,}{Pillepich
  et~al.}{2018}]{Pillepich2018}
Pillepich A.,  et~al., 2018, \mn@doi [\mnras] {10.1093/mnras/stx3112}, 475, 648

\bibitem[\protect\citeauthoryear{{Planck Collaboration} et~al.,}{{Planck
  Collaboration} et~al.}{2016}]{Planck2016}
{Planck Collaboration} et~al., 2016, \mn@doi [\aap]
  {10.1051/0004-6361/201525830}, 594, A13

\bibitem[\protect\citeauthoryear{Potter, Stadel  \& Teyssier}{Potter
  et~al.}{2017}]{Potter2017}
Potter D.,  Stadel J.,   Teyssier R.,  2017, \mn@doi [Comput. Astrophys.
  Cosmol.] {10.1186/s40668-017-0021-1}, 4, 2

\bibitem[\protect\citeauthoryear{Probst \& Boulesteix}{Probst \&
  Boulesteix}{2018}]{Probst2018}
Probst P.,  Boulesteix A.-L.,  2018, J. Mach. Learn. Res., 18, 1

\bibitem[\protect\citeauthoryear{Raissi, Perdikaris  \& Karniadakis}{Raissi
  et~al.}{2019}]{Raissi2019}
Raissi M.,  Perdikaris P.,   Karniadakis G.,  2019, \mn@doi [J. Comput. Phys.]
  {10.1016/j.jcp.2018.10.045}, 378, 686

\bibitem[\protect\citeauthoryear{Reddick, Wechsler, Tinker  \&
  Behroozi}{Reddick et~al.}{2013}]{Reddick2013}
Reddick R.~M.,  Wechsler R.~H.,  Tinker J.~L.,   Behroozi P.~S.,  2013, \mn@doi
  [\apj] {10.1088/0004-637X/771/1/30}, 771, 30

\bibitem[\protect\citeauthoryear{Rees \& Ostriker}{Rees \&
  Ostriker}{1977}]{Rees1977}
Rees M.~J.,  Ostriker J.~P.,  1977, \mn@doi [\mnras] {10.1093/mnras/179.4.541},
  179, 541

\bibitem[\protect\citeauthoryear{Riebe et~al.,}{Riebe et~al.}{2013}]{Riebe2013}
Riebe K.,  et~al., 2013, \mn@doi [Astron. Nachr.] {10.1002/asna.201211900},
  334, 691

\bibitem[\protect\citeauthoryear{Rosenblatt}{Rosenblatt}{1956}]{Rosenblatt1956}
Rosenblatt M.,  1956, \mn@doi [Ann. Math. Stat.] {10.1214/aoms/1177728190}, 27,
  832

\bibitem[\protect\citeauthoryear{Saintonge et~al.,}{Saintonge
  et~al.}{2013}]{Saintonge2013}
Saintonge A.,  et~al., 2013, \mn@doi [\apj] {10.1088/0004-637X/778/1/2}, 778, 2

\bibitem[\protect\citeauthoryear{Schaye et~al.,}{Schaye
  et~al.}{2015}]{Schaye2015}
Schaye J.,  et~al., 2015, \mn@doi [\mnras] {10.1093/mnras/stu2058}, 446, 521

\bibitem[\protect\citeauthoryear{Shankar et~al.,}{Shankar
  et~al.}{2017}]{Shankar2017}
Shankar F.,  et~al., 2017, \mn@doi [\apj] {10.3847/1538-4357/aa66ce}, 840, 34

\bibitem[\protect\citeauthoryear{Somerville \& Dav\'{e}}{Somerville \&
  Dav\'{e}}{2015}]{Somerville2015}
Somerville R.~S.,  Dav\'{e} R.,  2015, \mn@doi [\araa]
  {10.1146/annurev-astro-082812-140951}, 53, 51

\bibitem[\protect\citeauthoryear{Somerville \& Primack}{Somerville \&
  Primack}{1999}]{Somerville1999}
Somerville R.~S.,  Primack J.~R.,  1999, \mn@doi [\mnras]
  {10.1046/j.1365-8711.1999.03032.x}, 310, 1087

\bibitem[\protect\citeauthoryear{Springel et~al.,}{Springel
  et~al.}{2005}]{Springel2005}
Springel V.,  et~al., 2005, \mn@doi [\nat] {10.1038/nature03597}, 435, 629

\bibitem[\protect\citeauthoryear{Stiskalek, Bartlett, Desmond  \&
  Anbajagane}{Stiskalek et~al.}{2022}]{Stiskalek2022}
Stiskalek R.,  Bartlett D.~J.,  Desmond H.,   Anbajagane D.,  2022, \mn@doi
  [\mnras] {10.1093/mnras/stac1609}, 514, 4026

\bibitem[\protect\citeauthoryear{Vogelsberger, Marinacci, Torrey  \&
  Puchwein}{Vogelsberger et~al.}{2020}]{Vogelsberger2020}
Vogelsberger M.,  Marinacci F.,  Torrey P.,   Puchwein E.,  2020, \mn@doi [Nat.
  Rev. Phys.] {10.1038/s42254-019-0127-2}, 2, 42

\bibitem[\protect\citeauthoryear{Wetzel \& White}{Wetzel \&
  White}{2010}]{Wetzel2010}
Wetzel A.~R.,  White M.,  2010, \mn@doi [Monthly Notices of the Royal
  Astronomical Society] {10.1111/j.1365-2966.2009.16191.x}, 403, 1072

\bibitem[\protect\citeauthoryear{Wu, Pan, Chen, Long, Zhang  \& Yu}{Wu
  et~al.}{2021}]{Wu2020}
Wu Z.,  Pan S.,  Chen F.,  Long G.,  Zhang C.,   Yu P.~S.,  2021, \mn@doi [IEEE
  Trans. Neural Netw. Learn. Syst.] {10.1109/TNNLS.2020.2978386}, 32, 4

\bibitem[\protect\citeauthoryear{Yang \& Perdikaris}{Yang \&
  Perdikaris}{2019}]{Yang2019}
Yang Y.,  Perdikaris P.,  2019, \mn@doi [Comput. Mech.]
  {10.1007/s00466-019-01718-y}, 64, 417

\bibitem[\protect\citeauthoryear{Yarotsky}{Yarotsky}{2017}]{Yarotsky2017}
Yarotsky D.,  2017, \mn@doi [Neural Netw.]
  {https://doi.org/10.1016/j.neunet.2017.07.002}, 94, 103

\bibitem[\protect\citeauthoryear{Yasin, Desmond, Devriendt  \& Slyz}{Yasin
  et~al.}{2022}]{Yasin2022}
Yasin T.,  Desmond H.,  Devriendt J.,   Slyz A.,  2022, preprint (\mn@eprint
  {arXiv} {2210.07230})

\bibitem[\protect\citeauthoryear{Zhou, Chawla, Jin  \& Williams}{Zhou
  et~al.}{2014}]{Zhou2014}
Zhou Z.,  Chawla N.~V.,  Jin Y.,   Williams G.~J.,  2014, \mn@doi [IEEE Comput.
  Intell. Mag.] {10.1109/MCI.2014.2350953}, 9, 62

\bibitem[\protect\citeauthoryear{Zhou et~al.,}{Zhou et~al.}{2020}]{Zhou2020}
Zhou J.,  et~al., 2020, \mn@doi [AI Open] {10.1016/j.aiopen.2021.01.001}, 1, 57

\bibitem[\protect\citeauthoryear{{de Andres}, Yepes, Sembolini,
  {Mart{\'i}nez-Mu{\~n}oz}, Cui, Robledo, Chuang  \& Rasia}{{de Andres}
  et~al.}{2022}]{deAndres2022}
{de Andres} D.,  Yepes G.,  Sembolini F.,  {Mart{\'i}nez-Mu{\~n}oz} G.,  Cui
  W.,  Robledo F.,  Chuang C.,   Rasia E.,  2022, \mn@doi [\mnras]
  {10.1093/mnras/stac3009}, 518, 111

\bibitem[\protect\citeauthoryear{{de Santi}, Rodrigues, {Montero-Dorta},
  Abramo, Tucci  \& Artale}{{de Santi} et~al.}{2022}]{deSanti2022}
{de Santi} N. S.~M.,  Rodrigues N. V.~N.,  {Montero-Dorta} A.~D.,  Abramo
  L.~R.,  Tucci B.,   Artale M.~C.,  2022, \mn@doi [\mnras]
  {10.1093/mnras/stac1469}, 514, 2463

\bibitem[\protect\citeauthoryear{{von Marttens} et~al.,}{{von Marttens}
  et~al.}{2022}]{vonMarttens2022}
{von Marttens} R.,  et~al., 2022, \mn@doi [\mnras] {10.1093/mnras/stac2449},
  516, 3924

\makeatother
\end{thebibliography}

\bsp	
\label{lastpage}
\end{document}